\documentclass[12pt]{iopart}
\usepackage{iopams}  

\usepackage{xcolor}
\usepackage{graphicx}

\usepackage{subfig}
\usepackage{float}

\newcommand*{\del}{\partial}
\newcommand*{\dd}{\mathrm{d}}

\newcommand*{\xib}{\bar\xi}
\newcommand*{\etab}{\bar\eta}

\newcommand*{\rhob}{\bar\rho}
\newcommand*{\sigmab}{\bar\sigma}
\newcommand*{\mub}{\bar\mu}

\begin{document}
\title[Can gravitational waves halt the expansion of the universe?]{Can gravitational waves halt the expansion of the universe?}

\author{J Frauendiener$^1$, J Hakata$^2$ and C Stevens$^1$}

\address{${}^1$Department of Mathematics and Statistics, University of Otago, Dunedin 9016, New Zealand \\
${}^2$Department of Mathematics, Rhodes University, Grahamstown 6139, South Africa}

\ead{joergf@maths.otago.ac.nz, j.hakata@ru.ac.za, cstevens@maths.otago.ac.nz}

\begin{abstract}
We numerically investigate the propagation of plane gravitational waves in the form of an initial boundary value problem with de Sitter initial data. The full non-linear Einstein equations with positive cosmological constant $\lambda$ are written in the Friedrich-Nagy gauge which yields a wellposed system. The propagation of a single wave and the collision of two with colinear polarization are studied and contrasted with their Minkowskian analogues. Unlike with $\lambda=0$, critical behaviours are found with $\lambda>0$ and are based on the relationship between the wave profile and $\lambda$. We find that choosing boundary data close to one of these bifurcations results in a ``false'' steady state which violates the constraints. Simulations containing (approximate) impulsive wave profiles are run and general features are discussed. Analytic results of Woodard and Tsamis \cite{tsamis2013pure,tsamis2014classical}, which describe how gravitational waves could affect an expansion rate at an initial instance of time, are explored and generalized to the entire space-time. Finally we put forward boundary conditions that, at least locally, slow down the expansion considerably for a time.
\end{abstract}
\submitto{\CQG}
\maketitle

\section{Introduction}
Space-times containing plane gravitational waves have seen extensive analytical study over the years and many closed form solutions, which necessarily assume certain symmetries or wave profiles, now exist and their properties are known (see \cite{griffiths2016colliding} for an excellent overview.) While there are a number of analytic solutions for the propagation and collision of waves assuming a vanishing cosmological constant \cite{brinkmann1923riemann, peres1959some, takeno1961mathematical,khan1971scattering, penrose1972geometry, nutku1977colliding}, the non-vanishing cosmological constant analogues pale in number, and there are no closed form solutions for colliding waves in this case.

Penrose's cut-and-paste method \cite{penrose1972geometry, penrose1968twistor}, which cuts Minkowski space-time along a null hyperplane, shunts one half along the same surface and then pastes the two halves back together gives rise to a space-time with one impulsive  gravitational wave (i.e. with a Dirac delta function wave profile.) This has been generalized to non-zero, constant curvature backgrounds \cite{podolsky1999nonexpanding, podolsky1999expanding, griffiths2000exact, podolsky2000collision, podolsky2002exact, podolsky2019cut} where the wave fronts are topologically spherical for $\lambda>0$ and hyperboloidal for $\lambda<0$. There do not exist however, closed form solutions to the full non-linear Einstein equations with $\lambda\neq0$ that contain gravitational waves with \emph{plane symmetric} wave fronts.

De Sitter space-time, the unique solution to the Einstein vacuum equations with constant positive scalar curvature, can be thought of as a model of a universe which is expanding at an accelerated rate from the positive $\lambda$ contribution. Quantum gravitational back-reaction on inflation \cite{tsamis1997quantum} allows for the creation of cosmic scale gravitational radiation, where, if one does not account for their creation, can be modelled completely classically through gravitational perturbations of de Sitter space-time \cite{tsamis2013pure,tsamis2014classical}. It is theorized that such a background of radiation may weaken the expansion and even halt it completely. Analytical calculations have been done to explore this hypothesis by studying how an expansion parameter and its time derivative could be manipulated through such a field at an initial instance of time. The question of what happens away from this surface remains unanswered, and attempting to answer this in the full non-linear regime analytically would be very difficult, if not impossible.

In this paper, we numerically evolve the Einstein vacuum equations with positive cosmological constant in plane symmetry with the goal of shedding light on the above topics. To do so, we implement an initial boundary value problem following Friedrich and Nagy \cite{friedrich1999initial}, which is wellposed, and allows us to generate gravitational perturbations through boundary conditions rather than solving the constraints. This framework has already been implemented and numerically validated in previous work \cite{frauendiener2014numerical} for $\lambda=0$. We generalize this to an arbitrary cosmological constant as well as the inclusion of matter terms through components of $\Phi_{ab} = -(1/2)R_{ab} + (1/8)Rg_{ab}$ and scalar curvature $\Lambda = (1/24)R$ for completeness. We follow the conventions of Penrose and Rindler \cite{penrose1986spinors,penrose1988spinors} throughout.

\section{Review of plane gravitational waves with $\lambda=0$}
Here we briefly present the space-times of a single impulsive gravitational plane wave and the collision of two, which are colinearly polarized, with $\lambda=0$. This can be accomplished by summarizing the Khan-Penrose solution \cite{khan1971scattering}, which describes the latter.

\begin{figure}[H]
    \centering
    \includegraphics[width=0.4\linewidth]{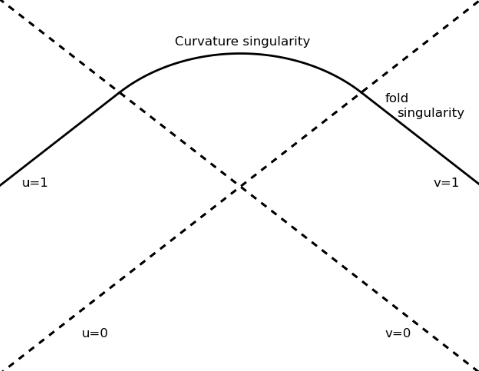}
    \caption{The structure of the Khan-Penrose solution.}
    \label{fig:kpsoln}
\end{figure}

Fig.~\ref{fig:kpsoln} showcases the Khan-Penrose solution in null coordinates $u,\,v$, where the two spatial dimensions that span the planes are suppressed, so that each point represents a plane. Null curves are represented by lines with slope $\pm1$ and the impulsive waves are given by $\Psi_0 = \delta(v),\,\Psi_4=\delta(u)$, where $\delta$ is the Dirac delta function, so their path is given by the dashed lines. These lines split the space-time into four regions. The lower region is Minkowski space-time, the two side regions are space-times containing one propagating wave only, and the top region is the interaction region after scattering. All four regions can be represented by the single line element
\begin{eqnarray}\label{eq:kpsoln}
    \textrm{d}s^2 &= \frac{2(1 - p^2 - q^2)^{3/2}}{\sqrt{1 - p^2}\sqrt{1 - q^2}(pq + \sqrt{1 - p^2}\sqrt{1 - q^2})^2}\textrm{d}u\textrm{d}v \nonumber \\
    &\quad -(1 - p^2 - q^2)\Big{(}\frac{\sqrt{1 - p^2} + q}{\sqrt{1 - p^2} - q}\Big{)}\Big{(}\frac{\sqrt{1 - q^2} + p}{\sqrt{1 - q^2} - p}\Big{)}\textrm{d}x^2 \nonumber \\
    &\quad -(1 - p^2 - q^2)\Big{(}\frac{\sqrt{1 - p^2} - q}{\sqrt{1 - p^2} + q}\Big{)}\Big{(}\frac{\sqrt{1 - q^2} - p}{\sqrt{1 - q^2} + p}\Big{)}\textrm{d}y^2,
\end{eqnarray}
where $p := u\,\Theta(u)$ and $q := v\,\Theta(v)$. The interaction region contains a spacelike curvature singularity on the surface $u^2 + v^2 = 1$ and can be seen as such due to the divergence of, for example, the Weyl invariant $I$. The region containing only the $\Psi_0$ wave is where $u<0$ and $v\geq0$ and the line element Eq.~\eref{eq:kpsoln} reduces to
\begin{equation}\label{eq:OneImpulsiveWave}
    \textrm{d}s^2 = 2\textrm{d}u\textrm{d}v - (1 + q)^2\textrm{d}x^2 - (1 - q)^2\textrm{d}y^2.
\end{equation}
This region and its $\Psi_4$ counterpart contain a \emph{fold singularity} along $v=1$ resp. $u=1$. As Eq.~\eref{eq:OneImpulsiveWave} can be transformed to Minkowski space-time by a coordinate transformation, one would think this is merely a coordinate singularity. However looking closer one sees that this is not the case as there does not exist a $C^1$ extension from this region to $v=1$ resp. $u=1$ \cite{matzner1984metaphysics}. Further, it is found that a certain projection of the $u=\;$constant, $v=\;$constant surfaces into Minkowski space in standard null coordinates converge at $v=1$. This has the consequence, which is discussed in more detail in Sec.~\ref{sec:AnalysisOfSingleWave}, that the spin-coefficients $\rho$ and $\rho'$, which when positive, represent the converging of a null geodesic congruence along $l^a$ and $n^a$ respectively, diverge to positive infinity, showing an ever strengthening contraction of null rays in both null directions.

\section{The equations}
\subsection{General setup}\label{sec:general-setup}
We write the Einstein equations in the form of an IBVP following Friedrich and Nagy \cite{friedrich1999initial} with the additional imposition of a pair of commuting space-like Killing vectors that represent our plane symmetry. Further, we include matter coming from an energy momentum tensor $T_{ab}$. A detailed explanation of this process in vacuum with vanishing cosmological constant has been laid out in \cite{frauendiener2014numerical}. We only give a brief summary here, emphasising the differences when including a non-vanishing cosmological constant and matter.

The Einstein equations take the form
\begin{equation}
    \Phi_{ab} + (3\Lambda - \frac12\lambda)g_{ab} = 4\pi T_{ab},
\end{equation}
where 
\begin{equation}
    R_{ab} = 6\Lambda g_{ab} - 2\Phi_{ab},
\end{equation}
and $\Lambda$ and $\Phi_{ab}$ correspond to the trace and tracefree part of the Ricci tensor $\Phi_{ab}$ and $\lambda$ is the cosmological constant.

To start setting up our gauge, we first assume our space-time can be foliated by planes. We then define the coordinates $t,z$ for time and the direction of wave propagation respectively, both being constant within the planes. Using the holonomic basis we define the null tetrad
\begin{eqnarray}
    l^a &= \frac{1}{\sqrt{2}}\Big{(}(1+B)(\del_t)^a + A(\del_z)^a\Big{)},\\
    n^a &= \frac{1}{\sqrt{2}}\Big{(}(1-B)(\del_t)^a - A(\del_z)^a\Big{)},\\[4pt]
    m^a &= \xi(\del_x)^a+\eta(\del_y)^a,
\end{eqnarray}
where $A,B,\xi,\eta$ are functions of $(t,z)$ only. This leads to the metric
\begin{equation}
    g = \dd t^2 - 2 \frac{B}{A}\, \dd t\dd z - \frac{1-B^2}{A^2}\, \dd z^2 + \frac2{(\xi\etab - \xib\eta)^2} \left(\eta\,\dd x - \xi\,\dd y\right)\left(\etab\,\dd x - \xib\,\dd y\right).
\end{equation}
To obtain equations for the metric functions and find algebraic relations for the spin-coefficients (due to the plane symmetry assumption) we apply the commutator equations (see \cite{penrose1986spinors} Eq. (4.11.11)) to the coordinates. To obtain equations for the spin-coefficients we use the curvature equations (see \cite{penrose1986spinors} Eq. (4.11.12)). To obtain equations and algebraic relations for the components of the Weyl tensor $C_{abcd}$, $\Phi_{ab}$ and $\Lambda$, we use the equations coming from the Bianchi identity (see \cite{penrose1986spinors} Eqs (4.12.36-4.12.41)).

The algebraic conditions are found to be
\begin{eqnarray}
    \rho = \rhob,\quad \rho' = \rhob',\quad \kappa = \kappa' = \alpha = \beta = \tau = \tau' = 0, \\[4pt]
    \Psi_1 = \Psi_3 = 0,\quad \Psi_2 = \sigma\sigma' - \rho\rho' + \Lambda + \Phi_{11}, \\[4pt]
    \Phi_{01} = \Phi_{10} = \Phi_{12} = \Phi_{21} = 0.
\end{eqnarray}
Following Friedrich and Nagy, we set
\begin{equation}
    \epsilon = \frac12(\rho - \rho' + F - \mu),\qquad
    \gamma = \frac12(\rho - \rho' + F + \mu),
\end{equation}
where the free function $F = \chi + i f$ is a freely specifiable gauge source function and $\mu$ is taken as a system variable. $\chi$ is the mean extrinsic curvature of the $z=$ constant hypersurfaces and $f$ determines the rotation of the $m^a$ frame vector along $(\partial_t)^a$. 

The geometrical interpretation of the new variable $\mu$ can be explained in the gauge $F = \rho' - \rho$, which is the gauge used for most of our results and turns out to be the Gau\ss\; gauge. Although predisposed to develop caustics, an expanding universe, which we consider here, acts to counter this. The fact we are in the Gau\ss\; gauge can be seen immediately by noticing that the only non-vanishing component of the acceleration of the unit time-like vector $(\partial_t)^a$ along itself is proportional to 
\begin{equation}
    \gamma + \bar{\gamma} + \epsilon + \bar{\epsilon} = F + \bar{F} + 2(\rho - \rho') = 0    
\end{equation}
for this choice of $F$. The ``acceleration'' $z^a\nabla_az^b$ of the space-like unit vector $z^a := A(\partial_z)^a$ along itself is proportional to $\mu + \bar{\mu}$, which gives an interpretation for the real part of $\mu$. The imaginary part just corresponds to a phase change of $m^a$.

It is found that the equations for $\eta,\xi$ decouple from the others, and as they are superfluous to the results subsequently presented we do not include them in the system.

The evolution equations are 
    \numparts
        \begin{eqnarray}
            \sqrt2 \del_t A &= (\mu + \bar\mu)\,A, \label{ee:1}\\
            \sqrt2 \del_t B &= (2\rho - 2\rho' + F + \bar F) + (\mu + \bar\mu) B,\label{ee:2}\\
            \sqrt2 \del_t \rho &= 3\rho^2 + \sigma \sigmab + \rho(F + \bar F) + \Phi_{00} - \Phi_{11} - 3\Lambda,\label{ee:3}\\
            \sqrt2 \del_t \rho' &= 3\rho^{\prime2}  + \sigma' \sigmab' - \rho'(F + \bar F) - \Phi_{11} + \Phi_{22} - 3\Lambda,\label{ee:4}\\
            \sqrt2 \del_t \sigma &= 4\rho\sigma - \rho'\sigma + \rho\sigmab' + \sigma(3F - \bar F) + \Psi_0,\label{ee:5}\\
            \sqrt2 \del_t \sigma' &= 4\rho'\sigma' - \rho\sigma' +
            \rho'\sigmab - \sigma'(3 F - \bar F) +
            \Psi_4,\label{ee:6}\\
            \sqrt2 \del_t \mu &= \mu^2 + \mu\bar \mu - 3 (\rho - \rho')^2 +
            (\mu + \bar \mu) (\rho + \rho') - \sigma \bar\sigma -
            \sigma'\bar\sigma' + 2 \sigma\sigma'\nonumber\\ 
            & - (\rho - \rho')(\bar F + 3F) - F^2 - F \bar F - \sqrt2 A\del_z F - 
            \sqrt2 B\del_t F \nonumber \\
            & - \Phi_{00} + 2\Phi_{11} - \Phi_{22} - 6\Lambda, \label{ee:7}\\[5pt]
            &\hspace{-3.9cm}(1-B) \del_t \Psi_0 - A \del_z \Psi_0 = \sqrt2 \left((2\rho - \rho' + 2 F + 2\mu)\Psi_0 + \sigma(3\Psi_2 + 2\Phi_{11}) + \sigmab'\Phi_{00}\right),\label{ee:8}\\
            &\hspace{-3.9cm}(1+B) \del_t \Psi_4 + A \del_z \Psi_4 = \sqrt2 \left((2\rho' - \rho - 2 F + 2\mu)\Psi_4 + 
            \sigma'(3\Psi_2 + 2\Phi_{11}) + \sigmab\Phi_{22}\right),\label{ee:9}
        \end{eqnarray}
    \endnumparts
while the constraints take the form
\numparts
    \begin{eqnarray}
        0=C_1 &:= \sqrt2 A\del_z\rho - (1 - 3 B) \rho^2 - (1 - B) \sigma\sigmab + \rho (\mu + \bar\mu + 2\rho')\nonumber \\
        &\quad + \rho B (F + \bar F) -(1-B)\Phi_{00} - (1+B)\Phi_{11} - 3(1+B)\Lambda,\label{ce:1}\\[4pt]
        0=C_2 &:= \sqrt2 A\del_z\rho' + (1 + 3 B) {\rho'}^2 + (1 + B) \sigma'\sigmab' - \rho' ( \mu + \bar\mu + 2\rho) \nonumber \\
        &\quad - \rho' B (F+\bar F) = (1-B)\Phi_{11} + (1+B)\Phi_{22} + 3(1-B)\Lambda,\label{ce:2}\\[4pt]
        0=C_3&:=\sqrt2 A\del_z\sigma + (1+B) \rho\sigmab' - 2 (1-2B)\rho\sigma + (1-B) \rho'\sigma  \nonumber\\ &\hskip8em + \sigma(3\mu - \bar\mu) + B\sigma(3F - \bar F) - (1-B) \Psi_0 ,\label{ce:3}\\
        0=C_4&:= \sqrt2 A\del_z\sigma' - (1-B) \rho'\sigmab + 2
        (1+2B)\rho'\sigma' - (1+B) \rho\sigma' \nonumber\\ &\hskip8em -
        \sigma'(3\mu - \bar\mu) - B\sigma'(3F - \bar F) + (1+B)
        \Psi_4. \label{ce:4}
    \end{eqnarray}
\endnumparts

To supplement the above, the divergence free condition on the energy-momentum tensor (equivalently the Bianchi identity, which are given in \cite{penrose1986spinors}, see Eq. 4.12.40) gives
\numparts
    \begin{eqnarray}
        &(1-B)\del_t\Phi_{00} + (1+B)(\del_t\Phi_{11} + 3\del_t\Lambda) \nonumber \\
        &= \sqrt{2}(2\rho + \mu + \mub + F + \bar F)\Phi_{00}
        + 4\sqrt{2}\rho\Phi_{11} \nonumber \\
        &\quad+ A(\del_z\Phi_{00} - \del_z\Phi_{11} - 3\del_z\Lambda),\label{dfe:1}\\
        &(1+B)\del_t\Phi_{22} + (1-B)(\del_t\Phi_11 + 3\del_t\Lambda) \nonumber \\
        &= \sqrt{2}(2\rho' + \mu + \mub - F - \bar F)\Phi_{22}
        + 4\sqrt{2}\rho'\Phi_{11}\nonumber \\
        &\quad+ A(\del_z\Phi_{11} - \del_z\Phi_{22} + 3\del_z\Lambda).\label{dfe:2}
    \end{eqnarray}
\endnumparts
Considering only the vacuum equations with cosmological constant, i.e. $\Phi_{ab}=0$, Eqs~\eref{dfe:1}--\eref{dfe:2} are identically satisfied and Eqs~\eref{ee:1}--\eref{ee:9}, Eqs~\eref{ce:1}--\eref{ce:4} comprise a closed system of equations, where the evolution equations are symmetric hyperbolic and the constraints propagate. When matter terms are present and one takes into account Eqs~\eref{dfe:1}--\eref{dfe:2}, it is still found that the above system is symmetric hyperbolic and the constraints propagate. The resulting subsidiary system is
\numparts
\begin{eqnarray}
    \sqrt{2}\del_tC_1 &= (6\rho + F + \bar F)C_1 + \sigmab C_3 + \sigma \overline{C_3}, \\
    \sqrt{2}\del_tC_2 &= (6\rho' - F - \bar F)C_2 + \sigmab' C_4 + \sigma' \overline{C_4}, \\
    \sqrt{2}\del_tC_3 &= (4\sigma + \sigmab')C_1 - \sigma C_2 + (4\rho - \rho' + 3F - \bar F)C_3 + \rho\overline{C_4}, \\
    \sqrt{2}\del_tC_4 &= (4\sigma' + \sigmab)C_2 - \sigma' C_1 + (4\rho' - \rho - 3F + \bar F)C_4 + \rho'\overline{C_3}.
\end{eqnarray}
\endnumparts
In order to close the system, one must in general couple it to equations describing the evolution of matter. There is a lot of freedom in this choice and it depends very much on the physical situation one wants to model. In general this choice will alter the principal part and, as a consequence, symmetric hyperbolicity and constraint propagation could be lost.

Two useful quantities are now introduced for monitoring the behaviour of the evolved space-time. First we note that the extrinsic curvature of our $t=\;$constant surfaces is $K_{ab}=-h_a^ch_b^d\nabla_ct_d$, where $h_{ab} = g_{ab} - t_at_b$ is the induced 3-metric on the surfaces and $t_a = (1-B^2)^{-1/2}(\textrm{d}t)_a$ is the unit conormal. We then define a local expansion parameter proportional to the mean extrinsic curvature $K_a{}^a$ as 
\begin{equation}
    \mathcal{H} := -\frac13K_a{}^a
    = \frac{\sqrt{2}(B^2-1)(B(F + \bar{F}) + \mu + \bar{\mu} + 2(\rho + \rho'))
    - 2 A \del_zB}{6(1-B^2)^{3/2}},
\end{equation}
which is used to monitor the expansion rate of the space-time along the time coordinate vector field. 
The Weyl scalar curvature invariants are useful tools for identifying whether a singularity is a curvature singularity. In the absence of matter and with our plane symmetry assumptions, the real part of $C_{abcd}C^{abcd}$ is the Weyl scalar curvature invariant
\begin{equation}\label{eq:KretschmannScalar}
    I := 2\Psi_0\Psi_4 + 6\Psi_2^2.
\end{equation}
We define the wave profile
\begin{equation}
  p(x) = \cases
        {
            32a\sin(bx)^8       & $\displaystyle0<x<\frac{\pi}{b}$ \cr
            0       & otherwise
        },
\end{equation}
where $b=35\pi/4$ so that the area of the profile is $a=\int_0^{\pi/b} p(x)\textrm{d}x$ and the amplitude is $32a$. We take $a$ as a measure of the strength of the wave. The boundary conditions for $\Psi_0$ and $\Psi_4$ will make use of $p(x)$ and are chosen in the subsequent sections. 

\subsection{De-Sitter space-time}
We investigate a variety of cases of plane gravitational waves propagating in de Sitter space-time (dS). The unperturbed metric in inflationary coordinates can be written \cite{hawking1973large}
\begin{equation}
    \dd s^2 = dt^2 - A_0^{-2}e^{2Ht}(\dd x^2 + \dd y^2 + \dd z^2), \qquad H^2=\lambda/3,\label{eq:dSLineElement}
\end{equation}
which covers half of the space-time and matches our setup for plane symmetry. This represents an expanding universe of the FLRW type. The appearance of $A_0 := A(0,z)$ is used to scale the spatial directions and will be useful later. It is useful to write dS in terms of null coordinates as
\begin{eqnarray}
    \dd s^2 &= e^{2Ht}\Big{(}2\dd u\,\dd v - (\dd x^2 + \dd y^2)\Big{)} \\
    &= 2\Big{(}\sqrt{2} - H(u + v + \sqrt{2})\Big{)}^{-2}\Big{(}2\dd u\,\dd v - (\dd x^2 + \dd y^2)\Big{)},
\end{eqnarray}
with transformations
\begin{eqnarray}
    u &= \frac{1}{\sqrt{2}}[H^{-1}(1 - e^{-Ht}) - A_0^{-1}(1+z)],\label{id:u}\quad\\
    v &= \frac{1}{\sqrt{2}}[H^{-1}(1 - e^{-Ht}) - A_0^{-1}(1-z)].\label{id:v}
\end{eqnarray}
The Minkowskian analogue of the above can be found in the limit $H\rightarrow0$.

In our formalism Eq.~\eref{eq:dSLineElement} gives the initial data
\begin{eqnarray}\label{eq:dSID}
    A = A_0,\quad \rho = \rho' = \mu = \pm\sqrt{\lambda/6},
\end{eqnarray}
with the remaining system variables, gauge quantities and matter terms vanishing. We will use the negative non-vanishing initial data, corresponding to a \emph{future expanding} universe, and set $\Phi_{ab}=0$.

We incorporate into the system null coordinates $u(t,z),v(t,z)$ which satisfy $l^a\nabla_au=0$ and $n^a\nabla_av=0$ respectively. Their initial and boundary data are fixed by Eqs~\eref{id:u}, \eref{id:v} so that when no wave is present we reproduce the same null coordinates as in Eq.~\eref{eq:dSLineElement} when $F$ is chosen appropriately. The above expressions for $u,v$ were chosen so that initially $u(0,-1) = 0 = v(0,1)$. Having $u,v$ available allows us to define the semi-invariant coordinates $(T,Z)$ by $T:=\sqrt{2}(v+u)$ and $Z:=\sqrt{2}(v-u)$ with which we can produce Penrose-Carter diagrams, i.e. diagrams where null curves are lines with slope $\pm1$. When in exact dS, as $t\rightarrow\infty$ we obtain $T\rightarrow2(H^{-1} - A_0^{-1})$ and $Z\rightarrow2zA_0^{-1}$.

\section{Numerical setup}
We utilize the Python package COFFEE \cite{doulis2019coffee}, which contains all the necessary functionality to perform a numerical evolution using the method of lines. We discretize the $z$-direction into equi-distant points in the interval $[-1,1]$ and approximate the $z$-derivative using Strand's finite difference stencil \cite{strand1994summation} which is fourth order in the interior, third order on the boundary and has the summation-by-parts property \cite{gustafsson1995time}. We march in time using the explicit fourth order Runge-Kutta scheme with a timestep determined by $\Delta t= c\,\Delta z$, where $\Delta z$ is the step size in the $z$-direction and $c$ is the CFL constant. Unless otherwise stated we take $c=0.5$. Boundary conditions are imposed using the Simultaneous Approximation Term (SAT) method \cite{carpenter1999stable} with $\tau=1$. This particular selection of numerical methods within COFFEE has proven to be numerically sound for a variety of different systems (see for example \cite{frauendiener2014numerical,beyer2017numerical}). In the subsequent situations, all constraints are verified to converge at the expected order everywhere.

\section{A single wave}\label{sec:SingleWave}
\subsection{An analytical view}\label{sec:AnalysisOfSingleWave}
Before analyzing the numerical results, it is worthwhile to perform a small analytic study of the propagation of one wave when either Minkowski or de Sitter initial data are taken.

Firstly, the evolution equations for $\rho$ and $\sigma$ (Eqs~\eref{ee:3} and \eref{ee:5}), which have a close relationship to Sachs' optical equations, give with vanishing $\Phi_{ab}$
\begin{eqnarray}
    \sqrt{2}\del_t\rho &= \rho(F + \bar{F}) + 3\rho^2 + \sigma\bar{\sigma} - 3\lambda,\qquad \\
    \sqrt{2}\del_t\sigma &= \sigma(3F - \bar{F} + 4\rho - \rho') + \rho\bar{\sigma}' + \Psi_0.
\end{eqnarray}
For the case of Minkowski initial data, which is obtained by setting $\lambda=0$ in the de Sitter initial data, and where we choose $F(t,z)=0$ to extend the exact gauge of dS to the whole space-time, we find the following: The introduction of a non-zero $\Psi_0$ on the right boundary causes $\sigma$ to become non-zero there. This in turn causes $\rho$ to become non-zero. As $\del_t\rho>0$, we find that $\rho$ will inevitably diverge. Further, one can see by looking at the evolution equations for the primed spin-coefficients, all primed spin-coefficients stay zero throughout the space-time, due to the forever zero $\Psi_4$. Further, this implies that $\Psi_2=0$ everywhere and thus the Weyl invariant $I$ given by Eq.~\eref{eq:KretschmannScalar} also remains zero everywhere. These are well known result for propagation of a single plane gravitational wave in Minkowski space-time, see \cite{griffiths2016colliding} for an overview (in a different gauge).

The case of expanding de Sitter initial data, with non-vanishing $\lambda$ and again choosing $F(t,z)=0$, is quite different. A non-zero $\Psi_0$ leads to a non-zero $\sigma$ as before, but now a non-zero $\sigma$ leads to a non-zero $\sigma'$ as well as a non-zero $\rho$. This non-zero $\sigma'$ then makes $\rho'$ and even $\Psi_4$ non-zero, implying the non-linear back-reaction effect is realized. This in turn leads to a non-zero Weyl invariant. The added complexity of the non-zero $\lambda$, which couples all system variables together in a complicated, non-linear way, stops us from concluding statements analogous to the Minkowski case as above, emphasising the need for numerics.

\subsection{Numerical analysis}
We now fix $\lambda=3$ and choose the boundary conditions to be
\begin{equation}
    \Psi_4(t,-1) = 0,\qquad \Psi_0(t,1) = p(v(t)),
\end{equation}
where $p(v)$ has the area of the wave packet as a parameter, and the change of area is realized by a change in amplitude. 
We perform evolutions with wave areas $a$ taking the values $1.67,\,1.6765,\,1.6769105,\,1.6769106,\,1.676912$, $1.67695,\,1.68$ for reasons that will become apparent shortly. In all these cases, once the wave has entered and subsequently left the computational domain, the space-time is fully excited in that all system variables have evolved away from their original values.

\begin{figure}[H]
    \centering
    \subfloat[\centering $\Psi_0$]
    {{\includegraphics[width=0.5\linewidth]{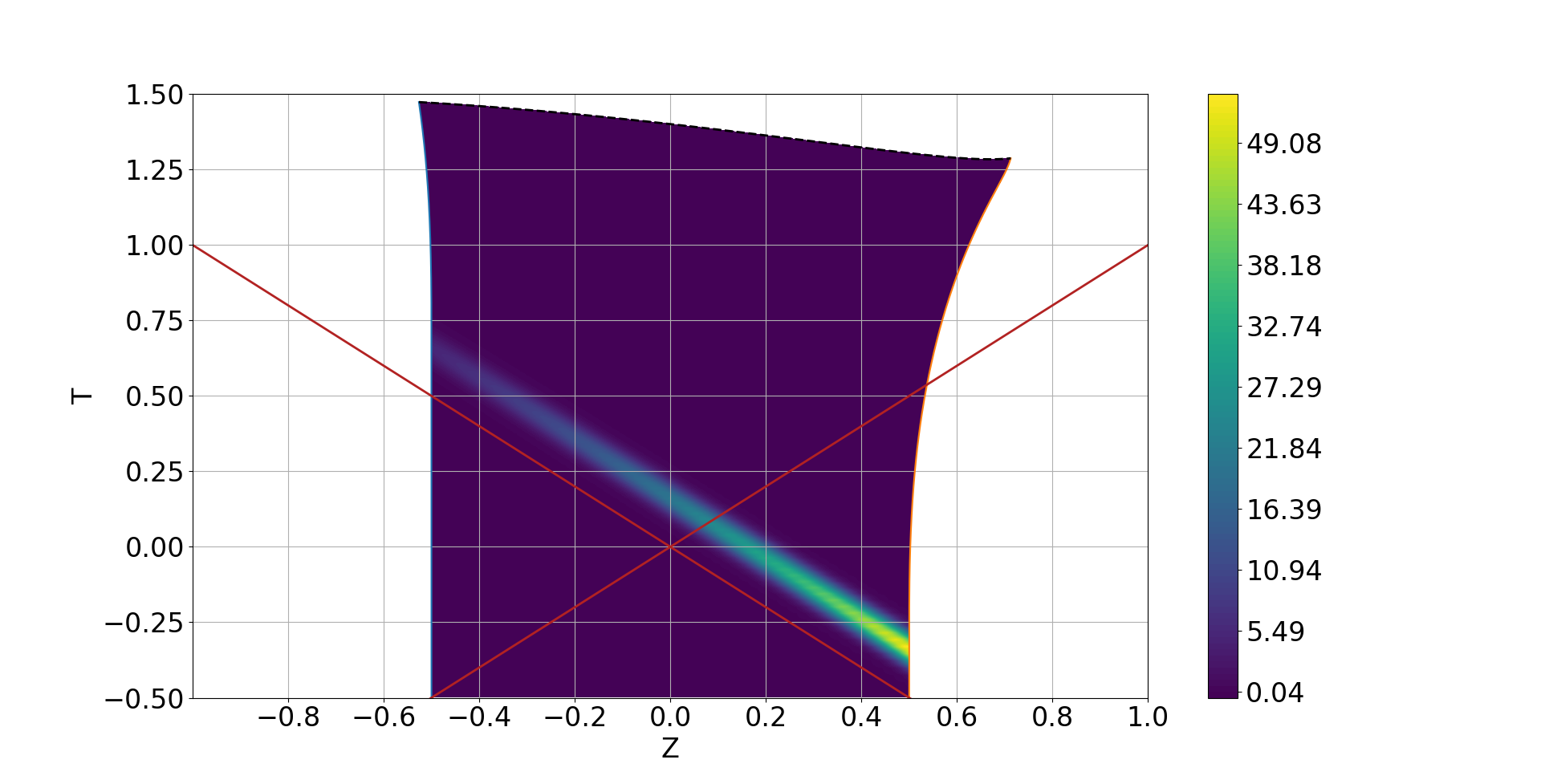}}}
    \qquad
    \subfloat[\centering $\mathcal{H}$]
    {{\includegraphics[width=0.5\linewidth]{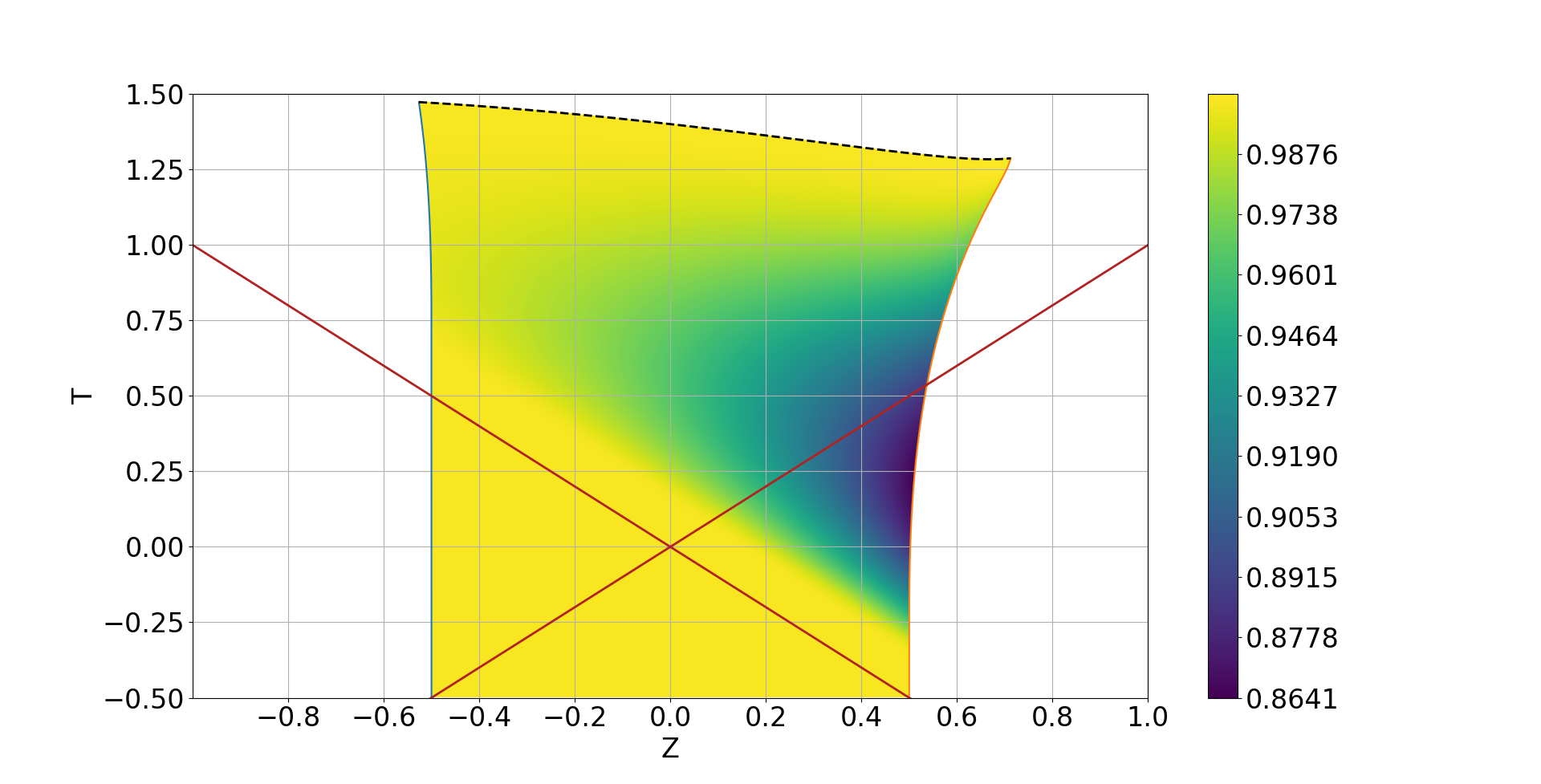}}}
    \caption{Contour plots of $\Psi_0$ and $\mathcal{H}$ plotted with respect to the semi-invariant $T$ and $Z$ coordinates where $a=1.67$. The dashed line represents the last timeslice and the crossed lines are $u=0$ and $v=0$.}
    \label{fig:Psi0andHOneWaveNoBlowup}
\end{figure}

For the case of four smallest values of $a$ we find that the space-time asymptotes back to the de Sitter space-time everywhere. This indicates that the wave has been wiped out by the accelerated expansion, already in stark contrast to the Minkowskian analogue where a future singularity is guaranteed. Fig.~\ref{fig:Psi0andHOneWaveNoBlowup} shows a contour plot of $\Psi_0$ and $\mathcal{H}$ over the entire space-time. It is clear that $\mathcal{H}$ decreases due to the addition of the gravitational wave, but then settles back down to its original value of one. The only remaining effect after the wave has passed is the time delay between different regions of space-time, such as the left and right boundaries.

To see how the representation of the null directions $l^a$ and $n^a$ in the coordinate basis change during the simulation, we look at the metric functions $A$ and $B$. It is seen that $A\rightarrow0$ as in the exact de Sitter case, representing the exponential expansion, and although initially $B$ increases to some value less than one, it asymptotes back to zero. Notably, the rate at which $A\rightarrow0$ and $B\rightarrow0$ causes the $\textrm{d}t\textrm{d}z$ metric coefficient to asymptote to a constant non-zero value and the $\textrm{d}z^2$ metric coefficient to diverge to positive infinity. The fact that $A$ and $B$ never actually reach zero implies our gauge remains regular.

\begin{figure}[H]
    \centering
    \includegraphics[width=0.5\linewidth]{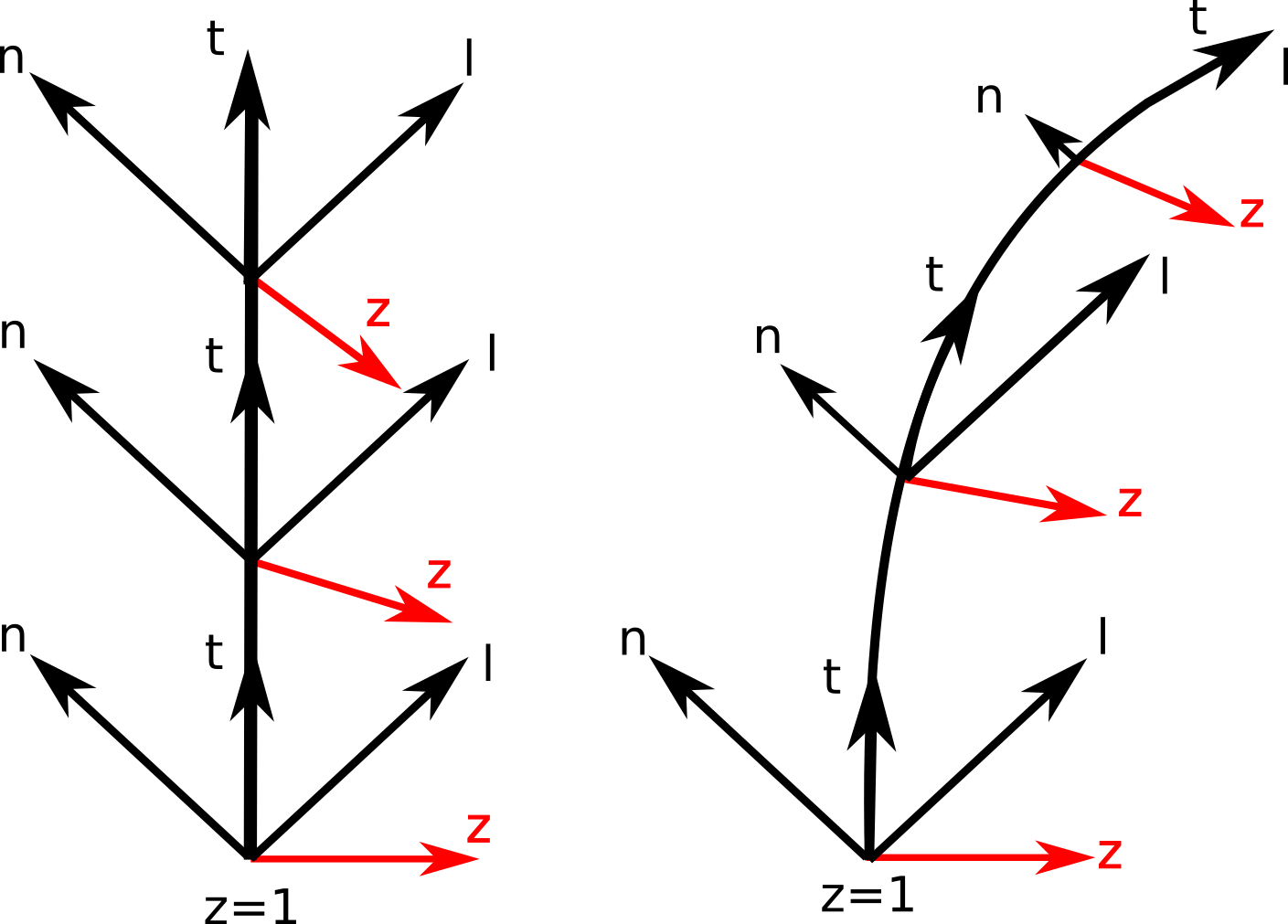}
    \caption{Two diagrams showcasing how the vectors $l^a$, $n^a, \partial^a_t$ and $\partial^a_z$ behave on the right boundary when a future singularity occurs.}
    \label{fig:RightBoundaryDiagram}
\end{figure}

For the four largest values of $a$ we find that the simulation crashes after some time due to $A\rightarrow0$ and $B\rightarrow1$ in finite time on the right boundary, the same as in the Minkowskian case. Fig.~\ref{fig:RightBoundaryDiagram} shows how this affects the relevant frame vectors there, where we note the relationships
\begin{equation}
    t^a := \partial_t^a = \frac{1}{\sqrt{2}}\Big{(}l^a + n^a\Big{)},\qquad
    z^a := A \partial_z^a = \frac{1}{\sqrt{2}}\Big{(}(1-B)l^a - (1+B)n^a\Big{)},
\end{equation}
where $t^a$ and $z^a$ are normalised. The left diagram is with respect to the $\{l^a,n^a\}$ null basis defined in the tangent space and exemplifies the fact that $t^a = \partial_t^a$ and is always normalised to one. It also showcases that the evolution of $z^a$ can cause trouble. This can be seen by noting that as $B\rightarrow1$ the $z=\;$constant surfaces become characteristic. The right diagram looks at another potential issue, this time in our $(T,Z)$-coordinates. In this case $t^a$ is no longer given by a vertical line, but $l^a$ and $n^a$ remain as lines with slope $\pm1$ from the definition of $u$ and $v$. The ``shrinking'' of the $n^a$ and the ``growing'' of $l^a$ is due to both coefficients of $n^a$ in the coordinate basis approaching zero, and enforces that $t^a$ is proportional to the sum of the two and that their normalisation conditions are maintained.

\begin{figure}[H]
    \centering
    \subfloat[\centering $I$]
    {{\includegraphics[width=0.5\linewidth]{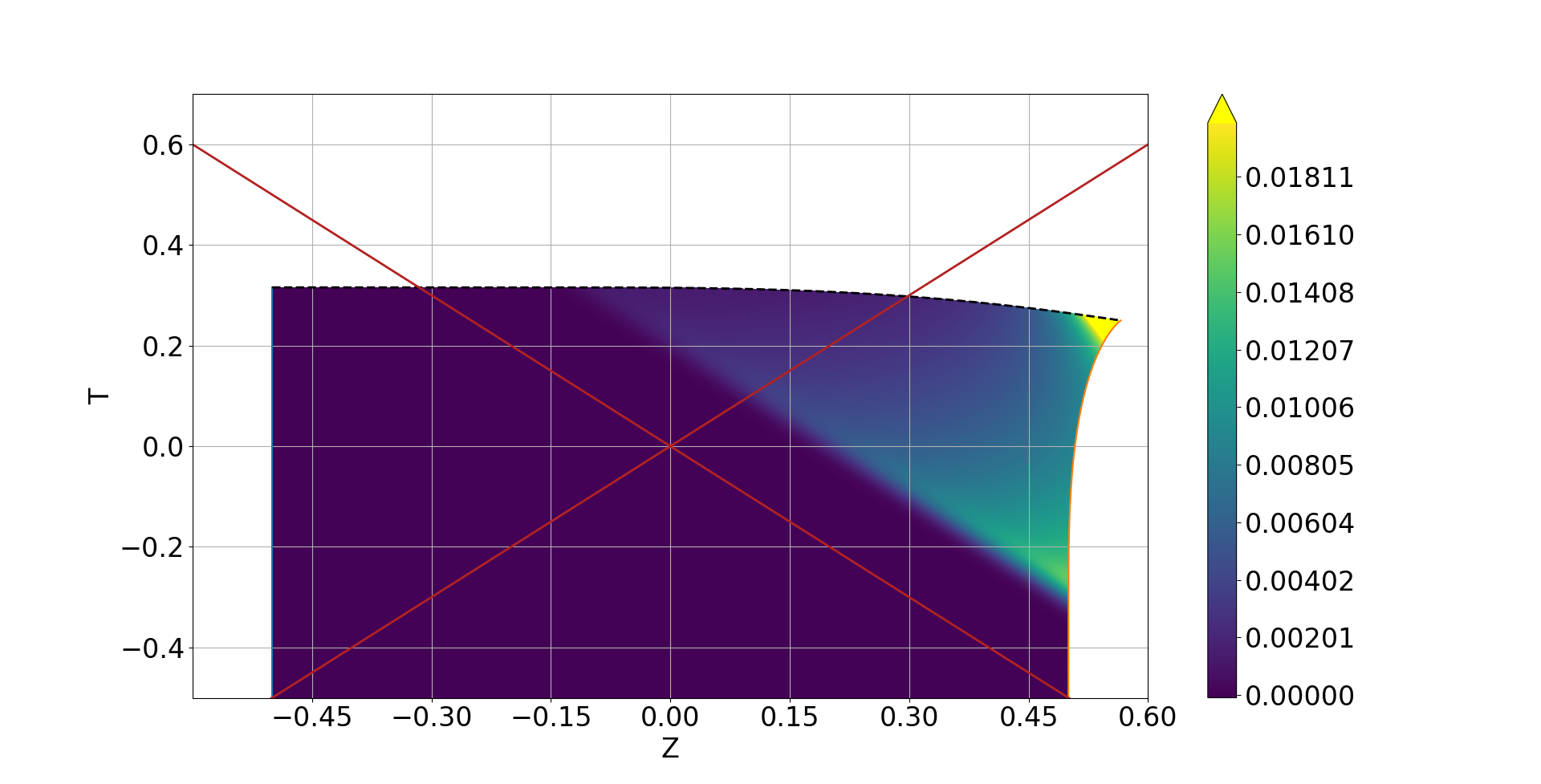}}}
    \qquad
    \subfloat[\centering $\mathcal{H}$]
    {{\includegraphics[width=0.5\linewidth]{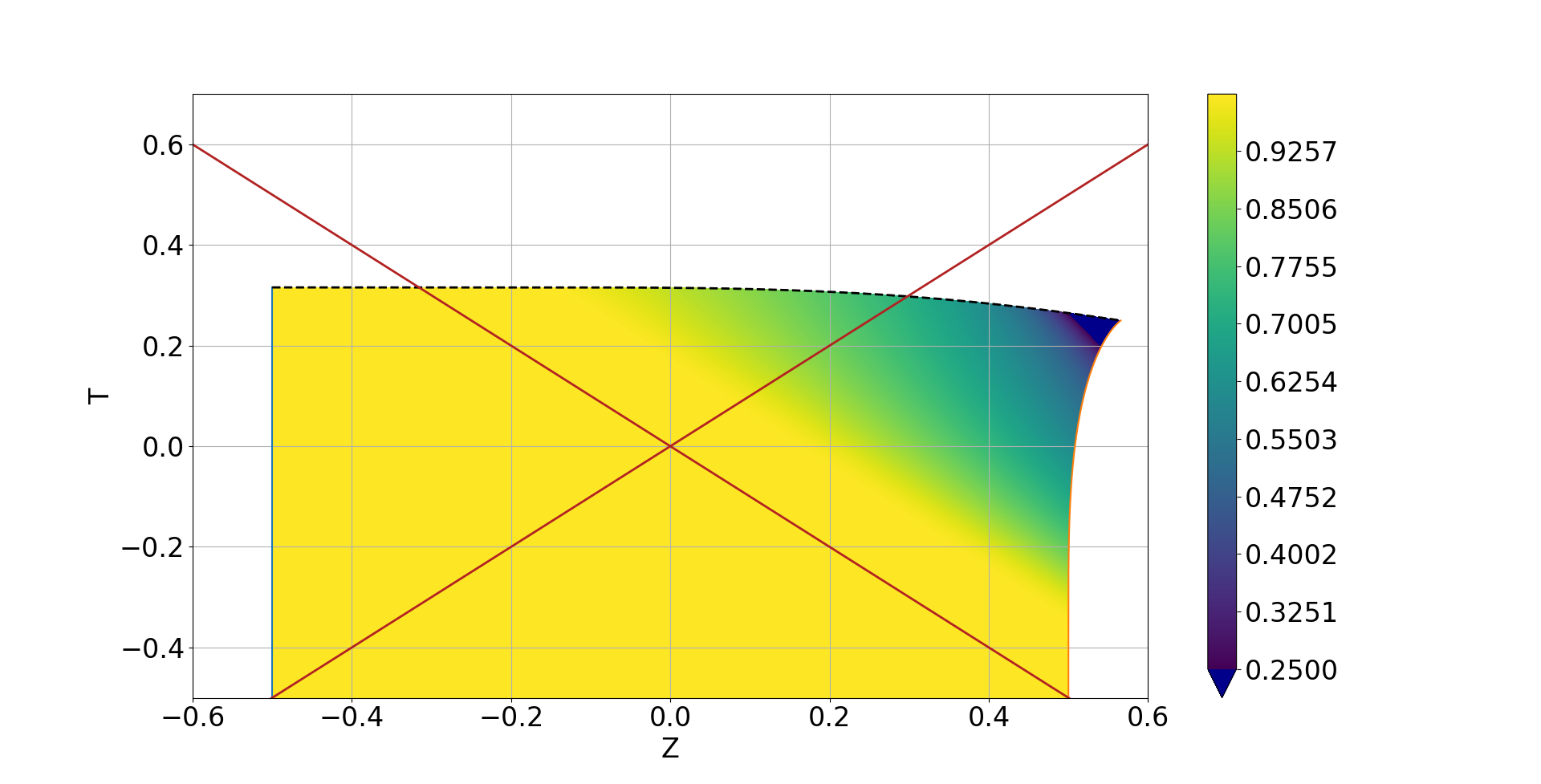}}}
    \caption{Contour plots of $I$ and $\mathcal{H}$ plotted with respect to the semi-invariant $T$ and $Z$ coordinates where $a=3$. The dashed line represents the last timeslice.}
    \label{fig:IandHOneWaveBlowup}
\end{figure}

This behaviour affects the expansion rate $\mathcal{H}$ and we find it decreases and actually diverges to $-\infty$ on the right boundary, as shown in Fig.~\ref{fig:IandHOneWaveBlowup}.

\begin{figure}[H]
    \centering
    \includegraphics[width=0.5\linewidth]{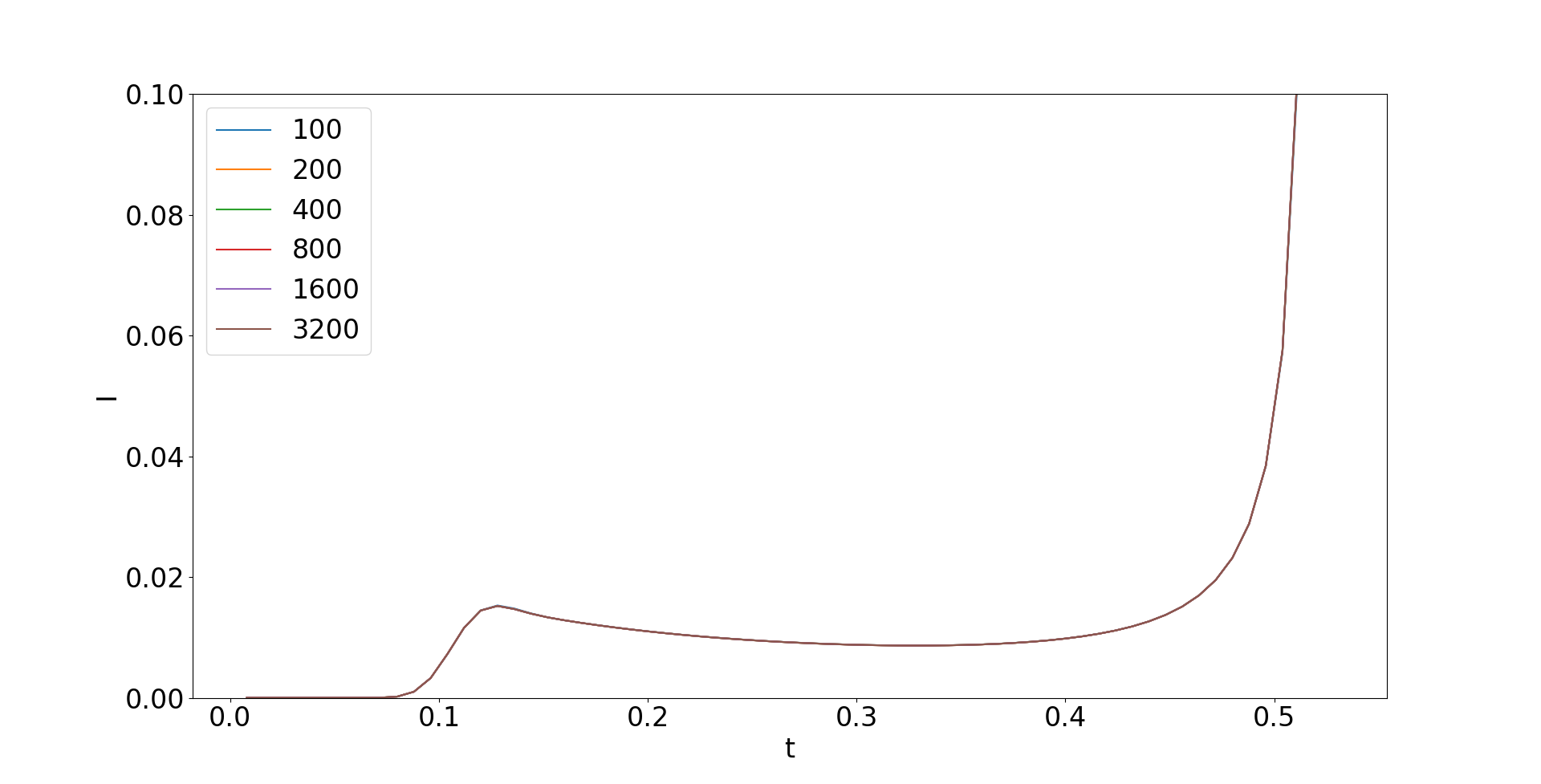}
    \caption{The Weyl invariant $I$ along the right boundary with $a=3$ for multiple $z$-resolutions which all fall within the same drawn curve.}
    \label{fig:IAlongRightBoundary}
\end{figure}

The features discussed above indicate that for these larger wave areas, the expansion rate of the space-time is not strong enough to overcome the contractivity of the wave, and a future singularity is formed. In the Minkowski case the analogue is a \emph{fold singularity} as discussed in Sec.~\ref{sec:AnalysisOfSingleWave}. In our de Sitter case, Fig.~\ref{fig:IandHOneWaveBlowup} and Fig.~\ref{fig:IAlongRightBoundary} show that the Weyl invariant $I$ is diverging on the right boundary (and similarly close to the right boundary), unlike the Minkowski case, and adds emphasis to the classification of a curvature singularity.

A final note is that changing the polarization of the wave, implemented by replacing $p(x)$ with $e^{i\phi}p(x)$ for some real constant $\phi$, does not affect the expansion rate or Weyl invariant as seen in Fig.~\ref{fig:Psi0andHOneWaveNoBlowup}, Fig.~\ref{fig:IandHOneWaveBlowup} and Fig.~\ref{fig:IAlongRightBoundary}.

\subsection{Critical behaviour}\label{sec:criticalbehaviour}
An obvious question has been raised: What is the critical behaviour when the ingoing wave has the critical wave area $a_c$ that separates these two distinct futures? One can obtain $a_c$ by using a simple binary search. For $\lambda=3$ this is found to be $1.67691055 < a_c < 1.67691056$.

\begin{figure}[H]
    \centering
    \subfloat[\centering $\rho$]
    {{\includegraphics[width=0.5\linewidth]{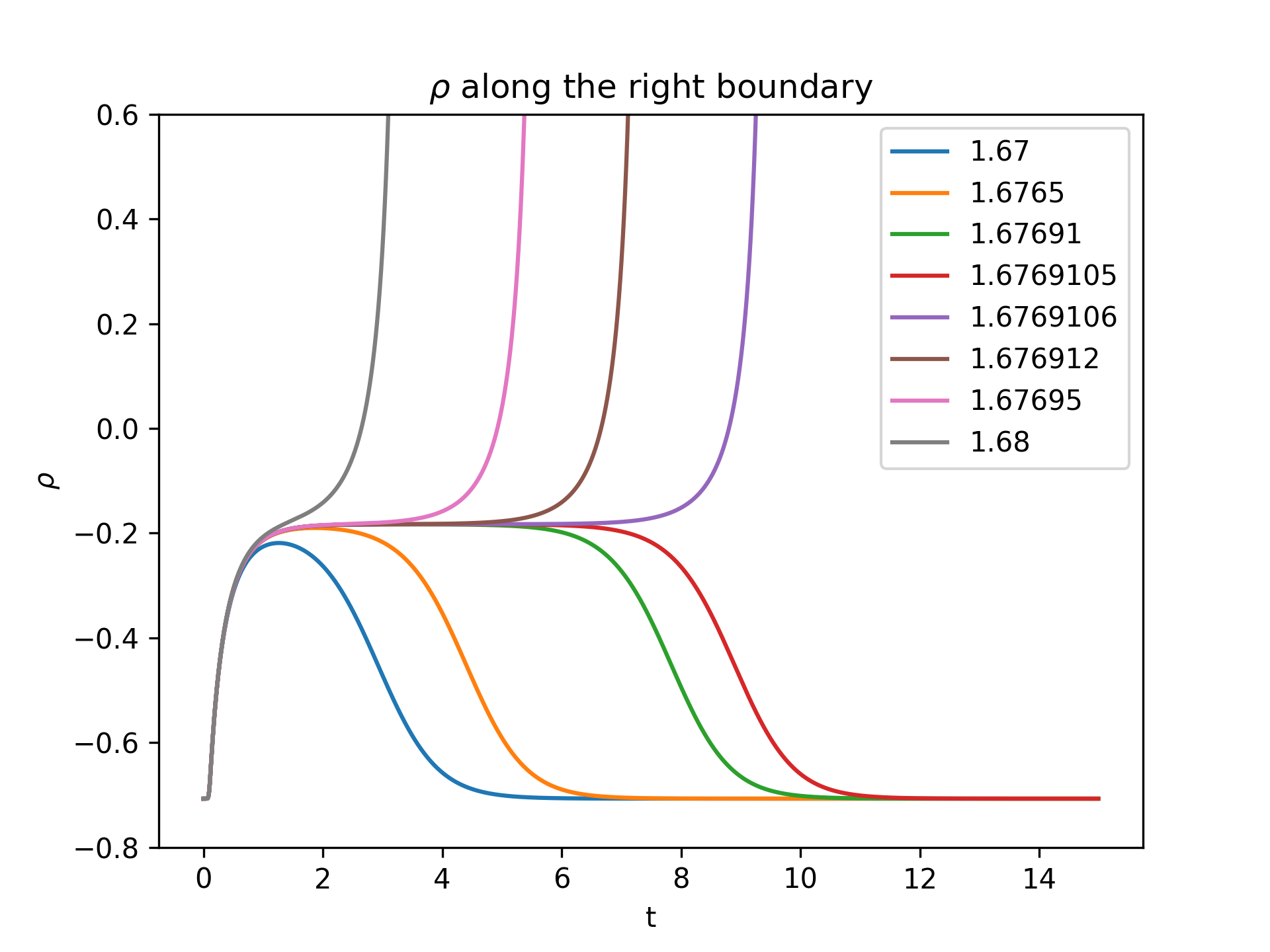}}}
    \qquad
    \subfloat[\centering $\sigma$]
    {{\includegraphics[width=0.5\linewidth]{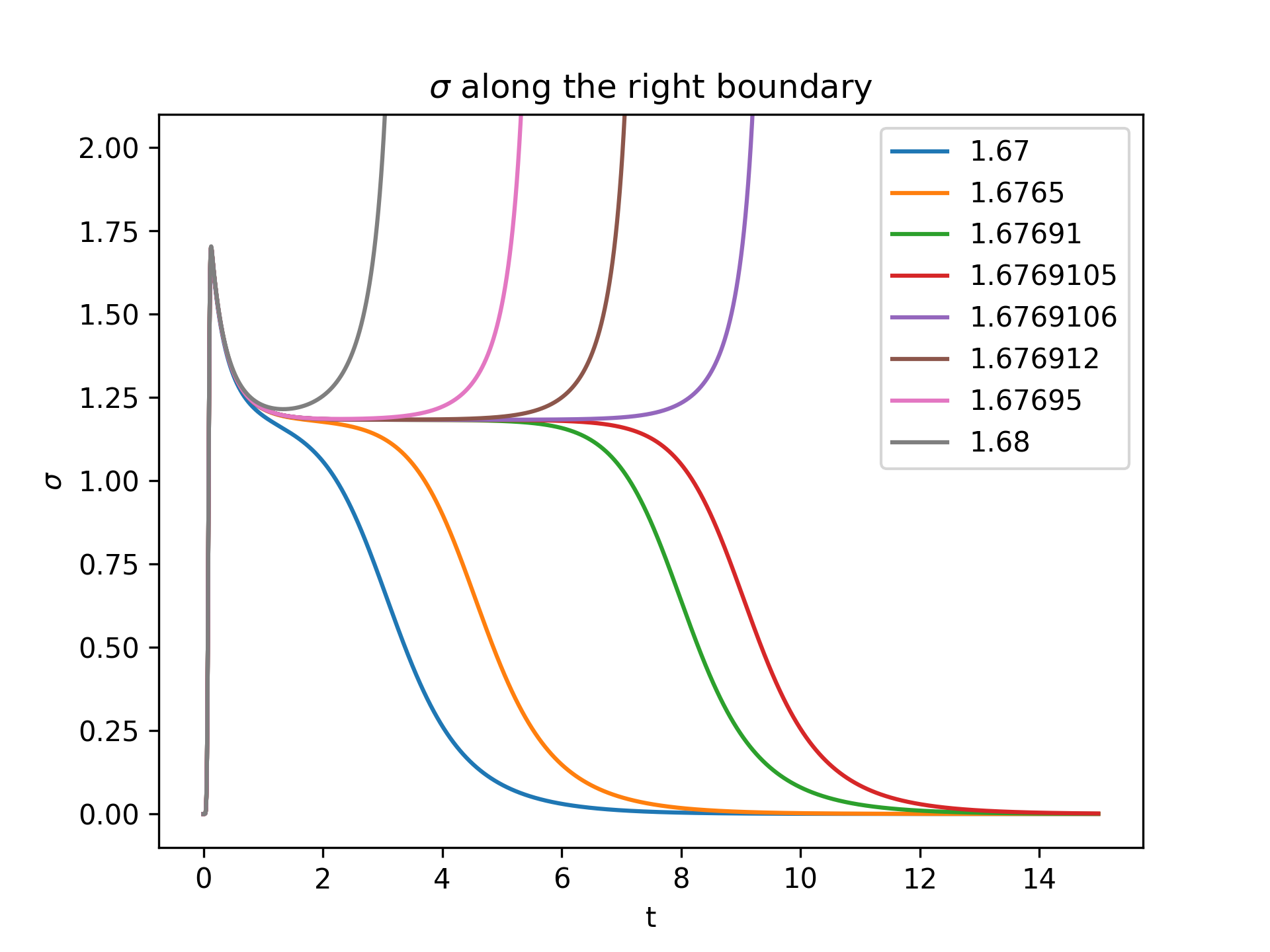}}}
    \caption{Plots of $\rho$ and $\sigma$ along the right boundary for different wave areas close to $a_c$. The curves corresponding to the first four values of $a$ from smallest to largest are the curves asymptoting back to their initial values from left to right. The curves corresponding to the larger four values of $a$ from smallest to largest are the curves which diverge from right to left.}
    \label{fig:rhosigma_AlongRightBoundary}
\end{figure}

Fig.~\ref{fig:rhosigma_AlongRightBoundary} shows $\rho$ and $\sigma$ along the right boundary for various wave areas close to $a_c$. It is clear that as $a\rightarrow a_c$ an interval appears where $\rho$ and $\sigma$ are constant in time and the interval becomes longer the closer $a$ is to $a_c$. This indicates that a special critical behaviour may exist.

\begin{figure}[H]
    \centering
    \includegraphics[width=0.65\linewidth]{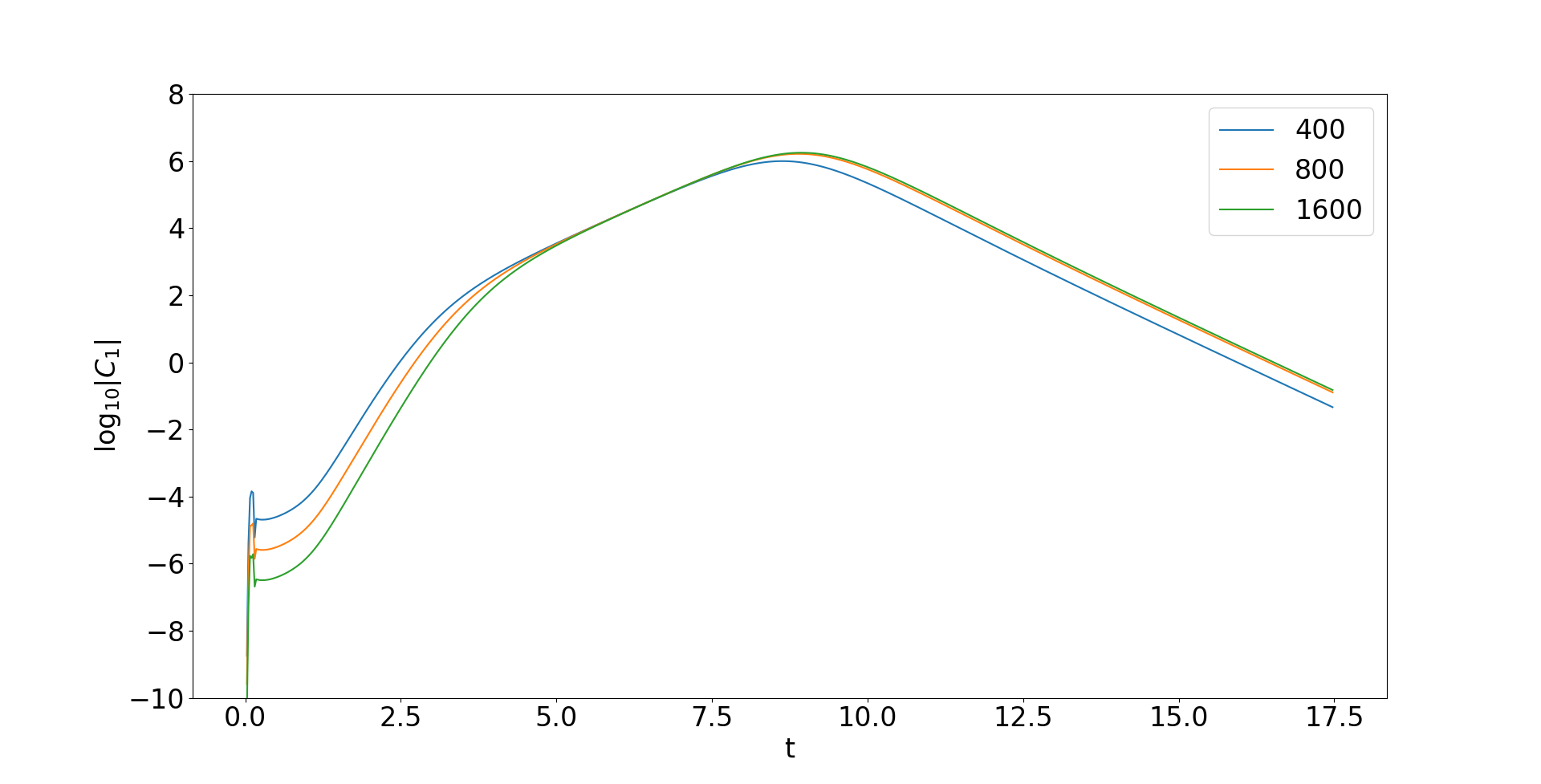}
    \caption{A convergence test for the constraint $C_1$ along the right boundary for the case of a single wave where $a=1.6769105$ and $\lambda=3$.}
    \label{fig:BifurcationConstraintViolation}
\end{figure}

All system variables except $A$ become constant in a finite $t$-interval on the boundary which becomes larger the closer to $a_c$ we take our wave area, and $\mu,\rho,\rho',\sigma,\sigma'$ take on values different than their initial ones. This implies a steady state solution, different from the de Sitter space-time. It turns out we can solve for unconstrained steady state solutions (but with $A$ a function of time) algebraically by setting all time derivatives except $A$ to zero in our evolution system, as well as taking $\Psi_0 = \Psi_4 = F = 0$. One of these solutions is found to match the values we see numerically. However, this exact solution \emph{does not} satisfy the constraint equations, and is thus a ``false'' steady state. This can be seen explicitly during our evolution, by noticing that the constraints do not converge and are wildy violated during this steady state period, see Fig.~\ref{fig:BifurcationConstraintViolation}. This is a consequence of our free evolution scheme, which by definition, is ``free'' from enforcing the constraints to be satisfied. It is found that the only free steady state solution (with $A$ varying in time) that also satisfies the constraints in the case of a positive cosmological constant with $\Psi_0 = \Psi_4 = F = 0$ is the de Sitter space-time.

The fact that no critical behaviour exists for this wave profile ansatz will be important when attempting to find a solution where the expansion is halted with gravitational radiation, and will be discussed in detail in Sec.~\ref{sec:SupressingExpansion}.

\subsection{An impulsive wave}\label{sec:OneImpulsiveWave}
Many analytical solutions describing gravitational waves in the literature have an impulsive wave profile, i.e. $\Psi_0 = \delta(v)$ where $v$ is a null coordinate, which is a consequence of the cut-and-paste method of Penrose \cite{penrose1972geometry}. An example is the propagation of a single impulsive gravitational plane wave with $\lambda=0$ given by Eq.~\eref{eq:OneImpulsiveWave}, where $\Psi_2 = 0 = \Psi_4$. To date, an exact solution for a single propagating plane gravitational wave with $\lambda>0$ has not been found. One cannot use Penrose's cut-and-paste method to find such a solution because this leads to wavefronts that are spherical or hyperboloidal when $\lambda>0$ or $\lambda<0$ respectively \cite{podolsky2019cut}. Thus, to try shed some light toward an analytic solution, we numerically evolve our system with $\lambda>0$ and with one ingoing wave, whose wave profile approximates the Dirac delta function. We set $\Psi_0(v,1) = q(v)$ where
\begin{equation}\label{eq:qBC}
    q(x) := \cases
          {
              a\sin(bx)^8       & $\displaystyle0<x<\frac{\pi}{b}$ \cr
              0       & otherwise
          },
\end{equation}
where $b = 35\pi a / 128$ and $q(x)$ has the property that $\displaystyle\lim_{a\rightarrow\infty}q(x)=\delta(x)$. We also change our gauge and fix $F$ by the condition that $\del_t B=0$. This matches the gauge of the exact solution given by Eq.~\eref{eq:OneImpulsiveWave} and yields $F = \rho' - \rho$. We choose $a=128,\,256,\,512,\,1024$, populate our spatial interval $z\in[-1,1]$ with $6401$ equi-distant points to accurately resolve these steep wave profiles and choose $\lambda=0.6$ and $\lambda=1.2$ to exemplify futures that do and do not have a singularity respectively.



\begin{figure}[H]
    \centering
    \subfloat[\centering $\Psi_0$]
    {{\includegraphics[width=0.33\linewidth]{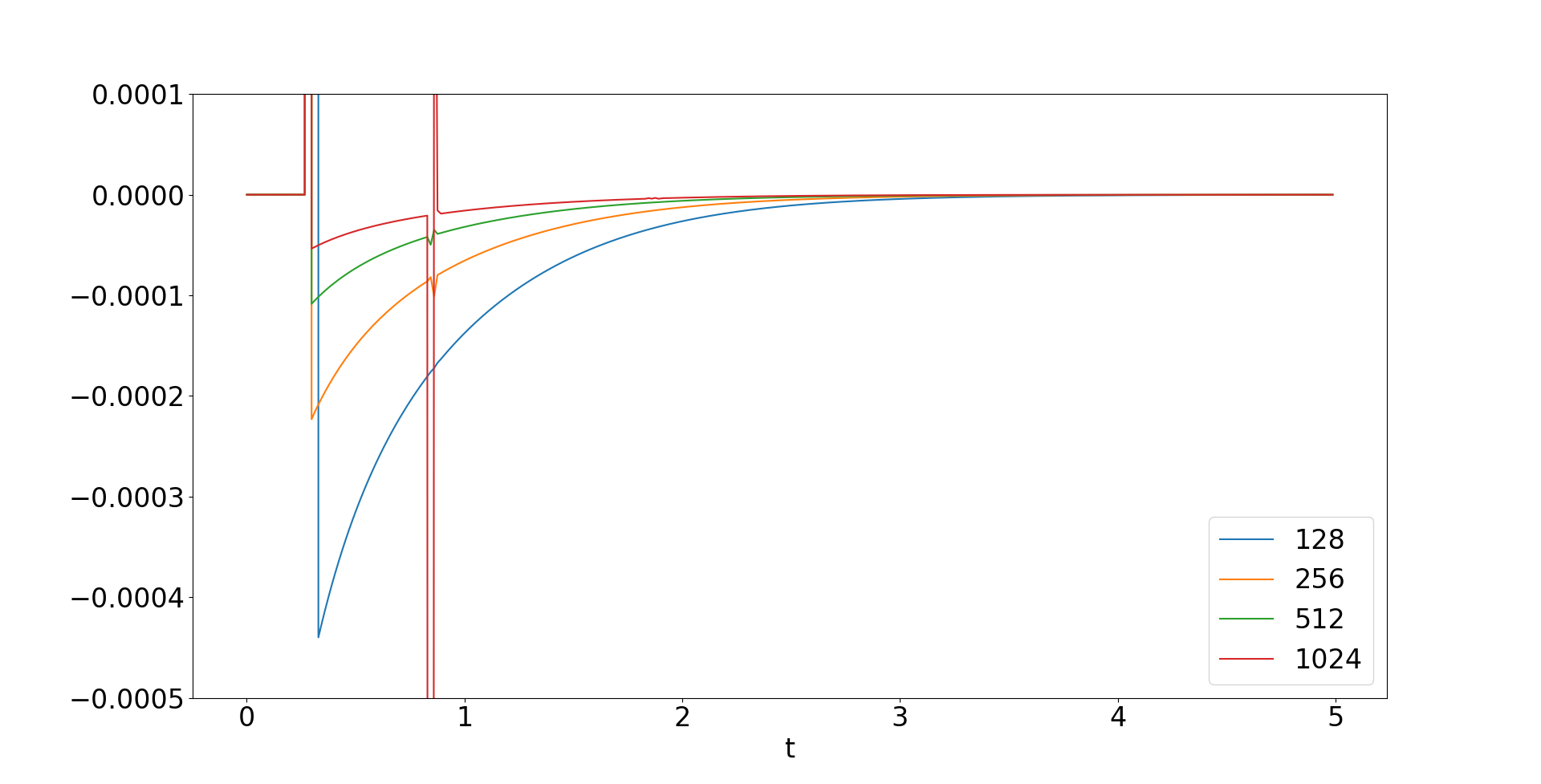}}}
    \qquad
    \subfloat[\centering $\Psi_2$]
    {{\includegraphics[width=0.33\linewidth]{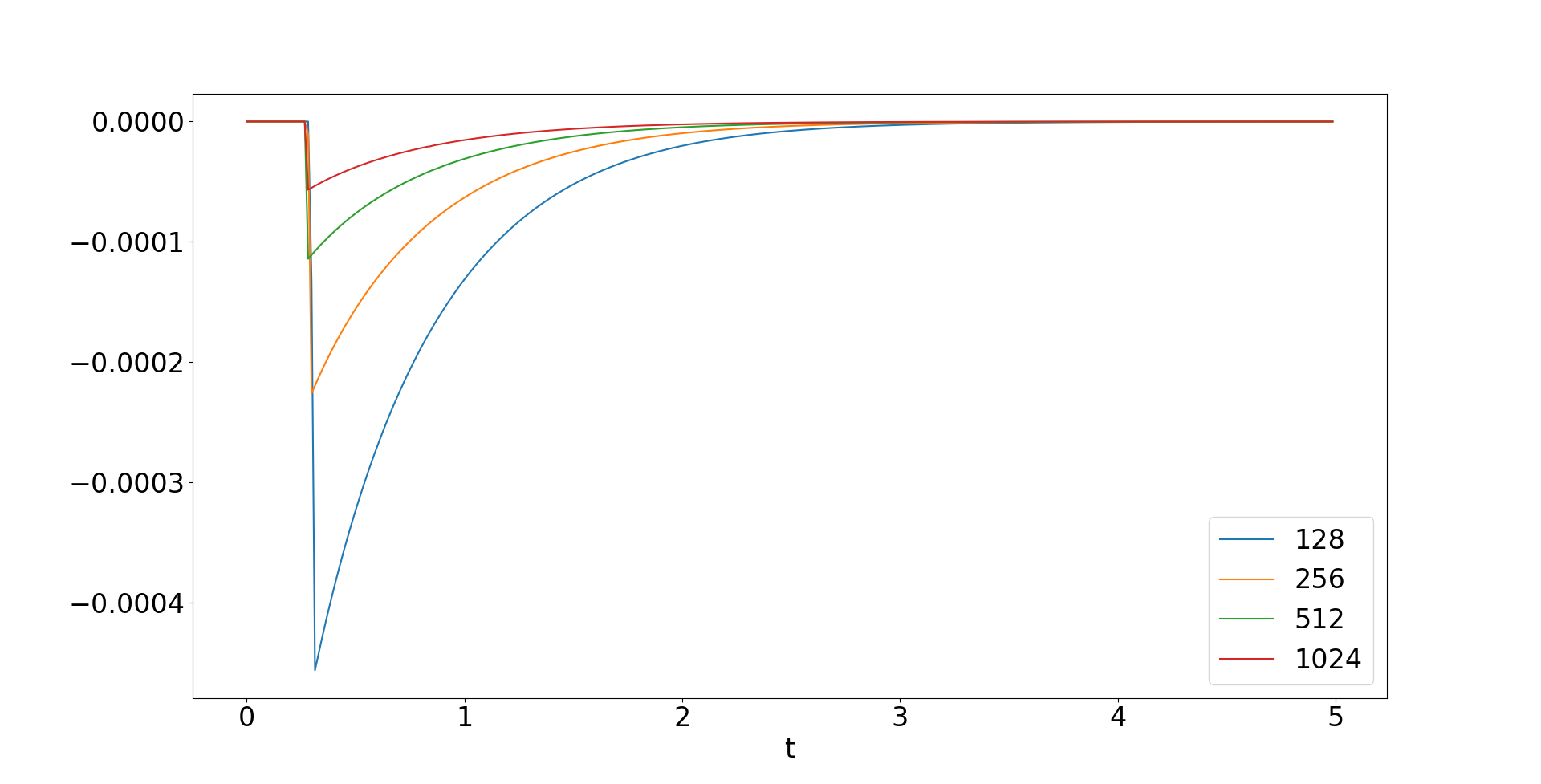}}}
    \qquad
    \subfloat[\centering $\Psi_4$]
    {{\includegraphics[width=0.33\linewidth]{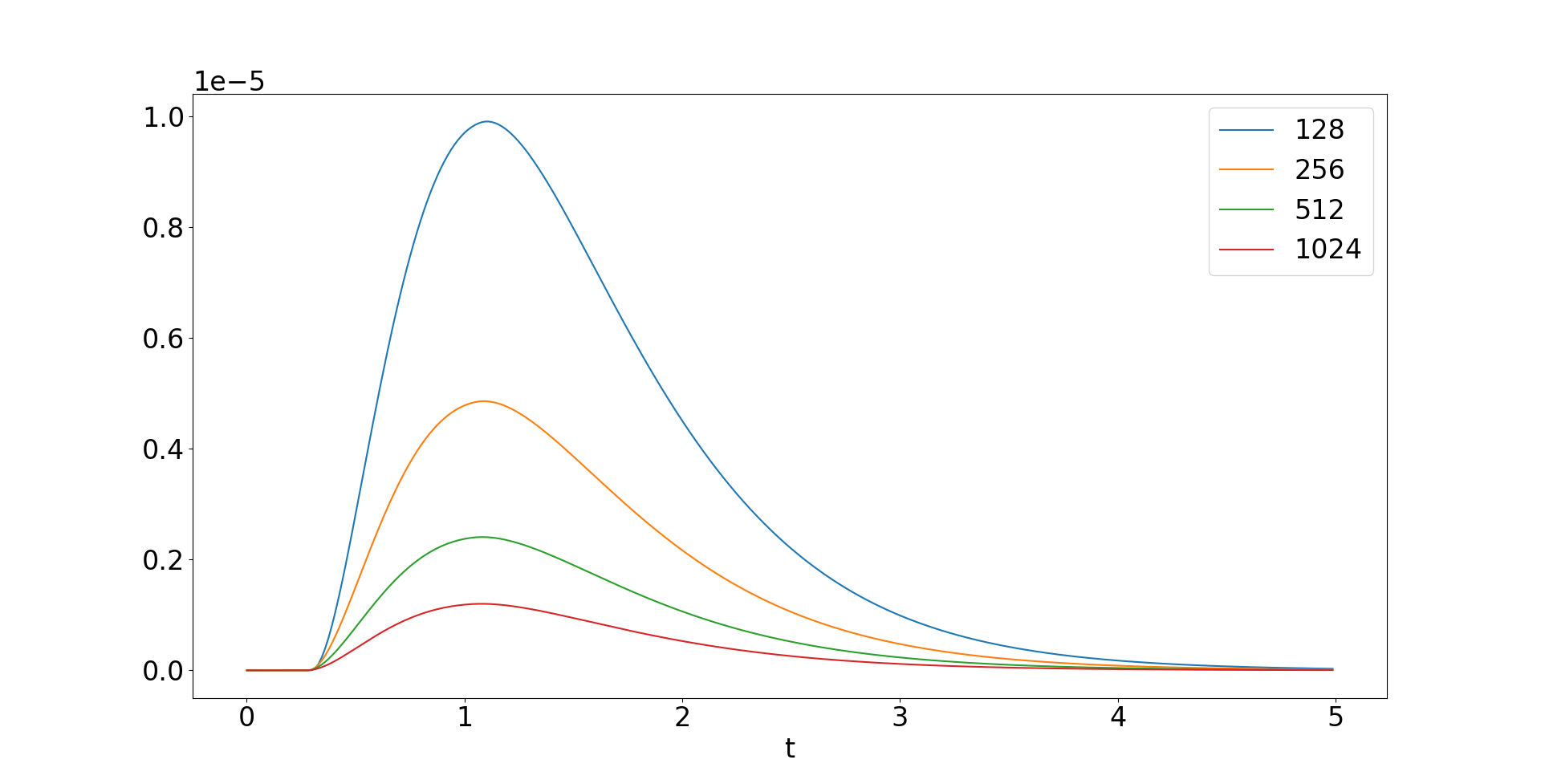}}}
    \caption{Plots along $z=0$ of $\Psi_0, \Psi_2$ and $\Psi_4$ as $a\rightarrow\infty$ with $\lambda=0.6$.}
    \label{fig:ImpulsiveWaveAlongz0Lambda0p2}
\end{figure}

For $\lambda=1.2$, to see the effect of the limit $a\rightarrow\infty$, Fig.~\ref{fig:ImpulsiveWaveAlongz0Lambda0p2} shows the Weyl components along $z=0$. These seem to indicate that in this limit, they all vanish for $v>0$ along $z=0$. By inspection it is clear that this happens along any $z=\;$constant curve once the wave has past and thus in the whole region $v>0$. Further, all system variables asymptote back to dS after the wave has past, and no singularity is formed. Note the numerical error in Fig.~\ref{fig:ImpulsiveWaveAlongz0Lambda0p2} (a) just before $t=1$. This is due to the steep wave profile and its interaction with the left boundary propagating back into the computational domain. This phenomenon is discussed in detail in Sec.~\ref{sec:CollidingImpulsiveWaves}.

\begin{figure}[H]
    \centering
    \subfloat[\centering $\Psi_2$ for $\lambda=0.6$]
    {{\includegraphics[width=0.5\linewidth]{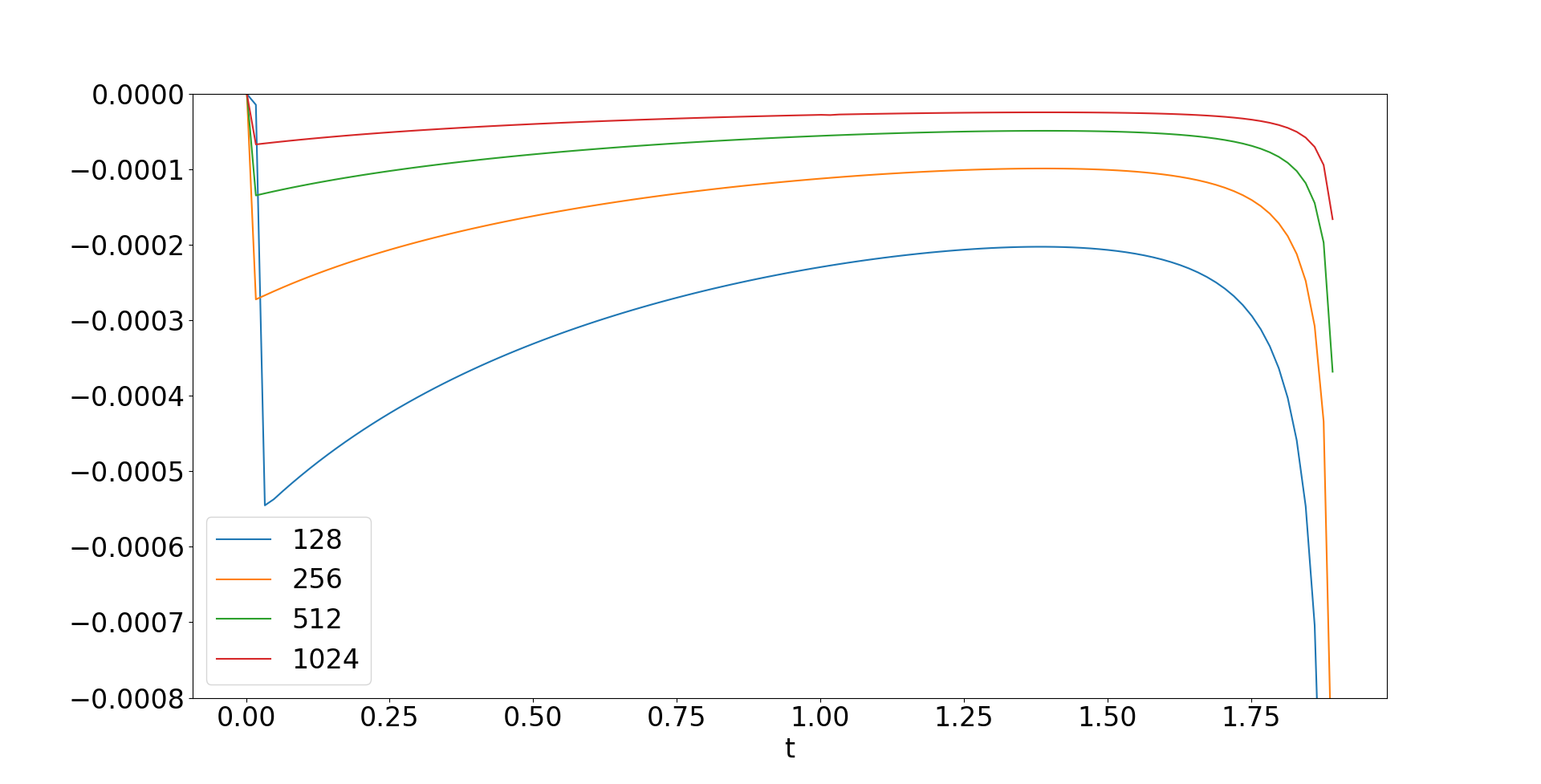}}}
    \qquad
    \subfloat[\centering $\Psi_2$ for $\lambda=1.2$]
    {{\includegraphics[width=0.5\linewidth]{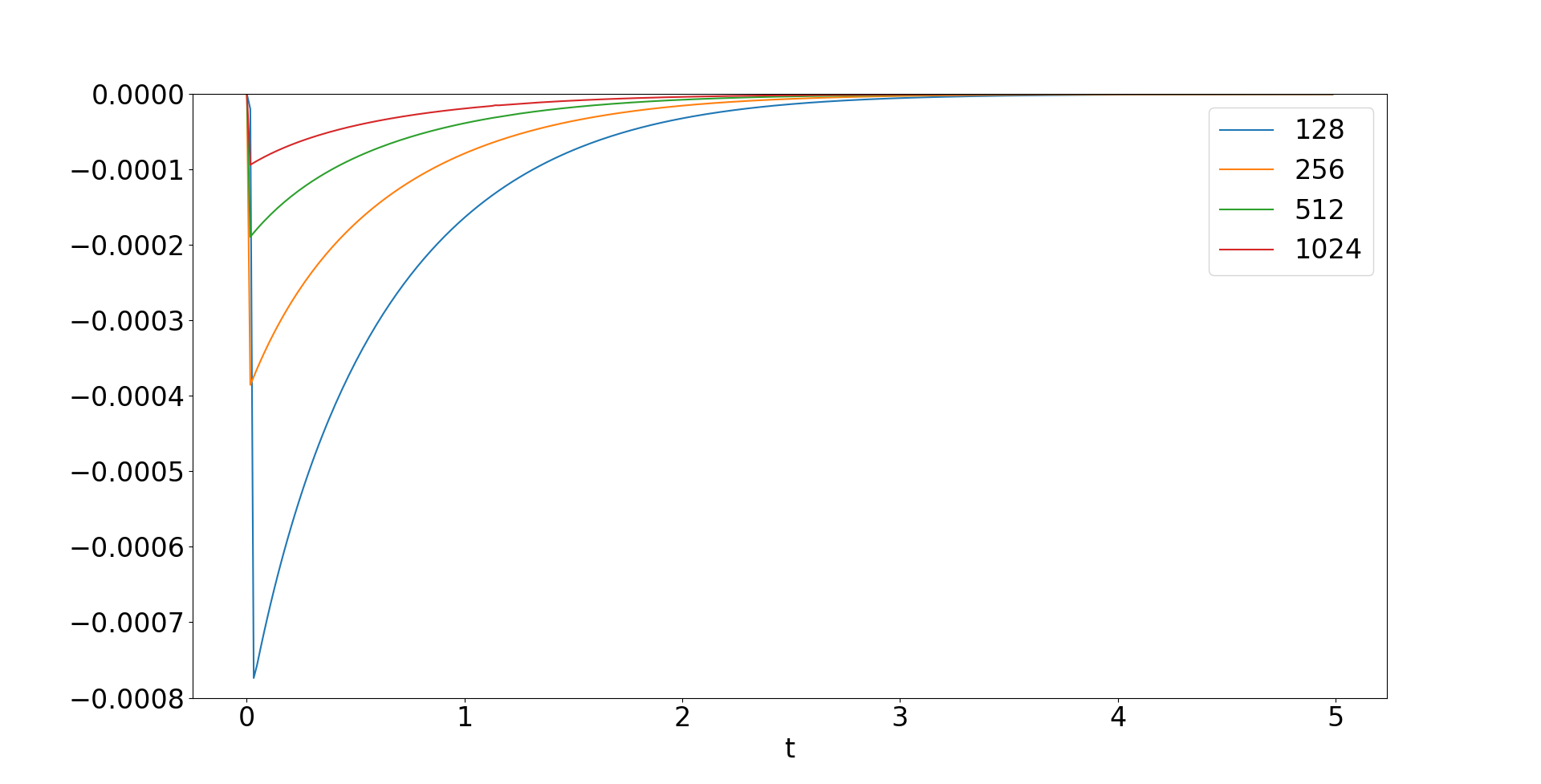}}}
    \\
    \subfloat[\centering $\Psi_4$ for $\lambda=0.6$]
    {{\includegraphics[width=0.5\linewidth]{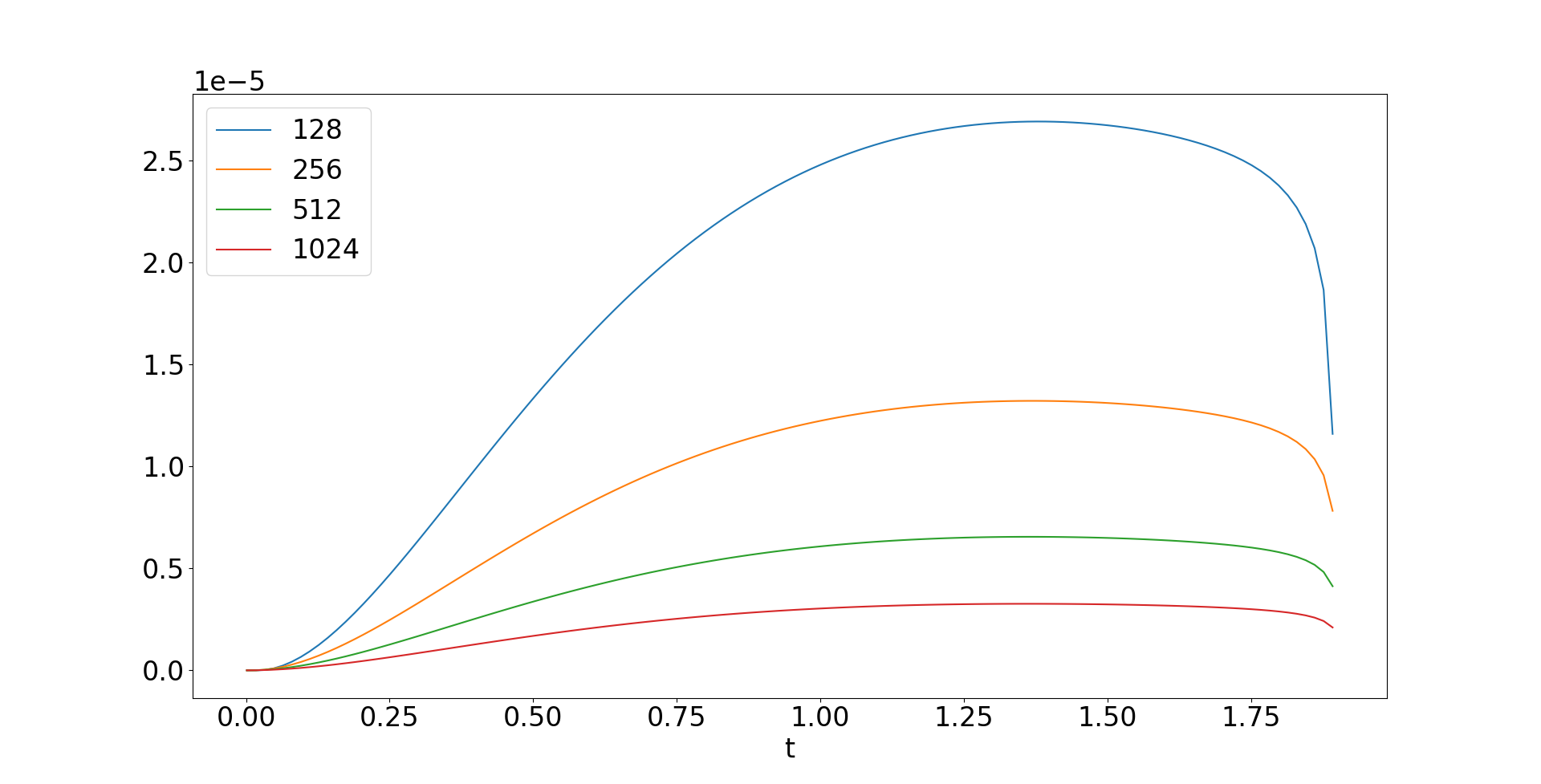}}}
    \qquad
    \subfloat[\centering $\Psi_4$ for $\lambda=1.2$]
    {{\includegraphics[width=0.5\linewidth]{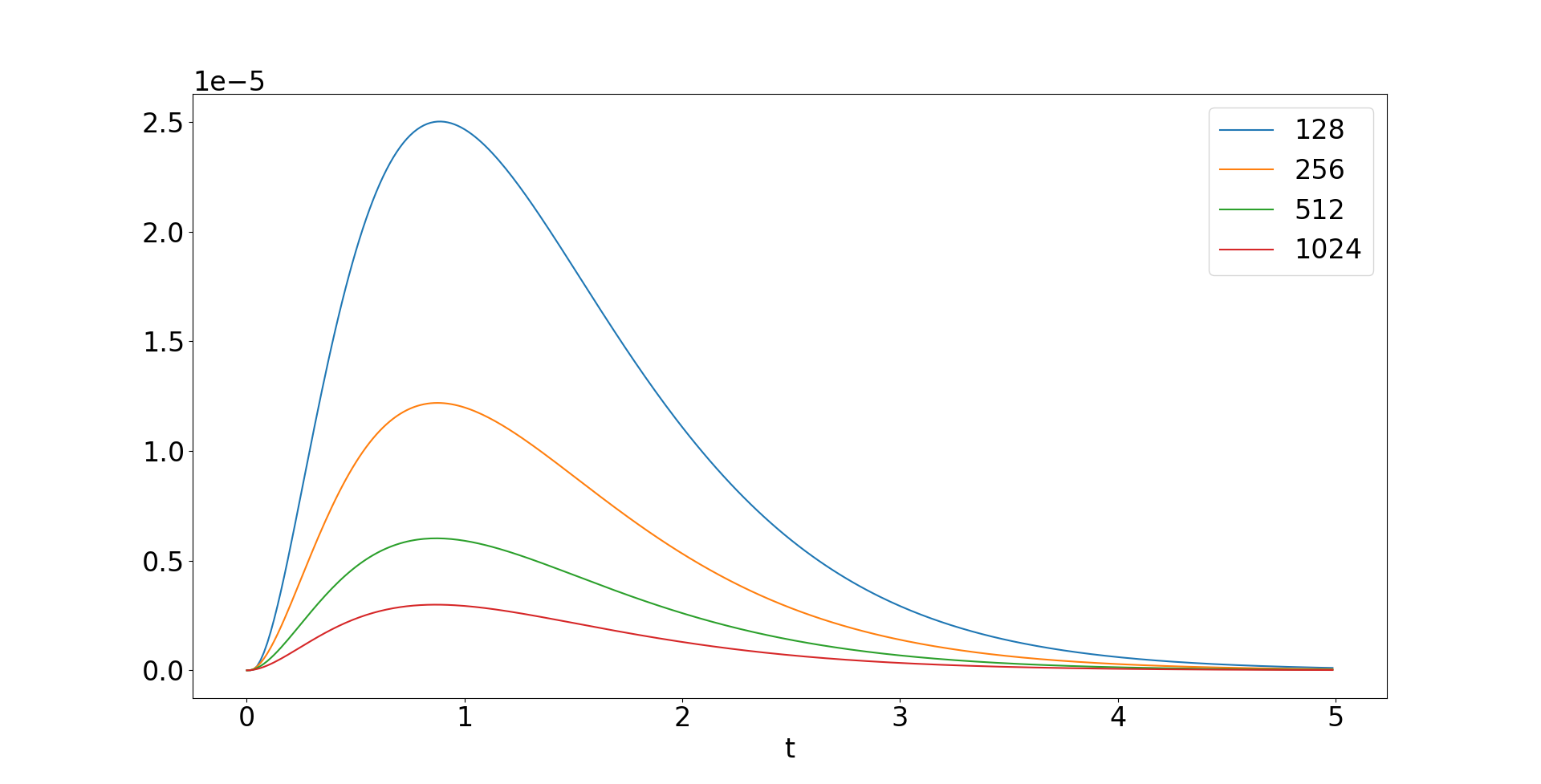}}}
    \caption{Plots along $z=1$ of $\Psi_2$ and $\Psi_4$ as $a\rightarrow\infty$ with $\lambda=0.6$ and $\lambda=1.2$.}
    \label{fig:ImpulsiveWaveAlongRightBoundaryLambda0p10p2}
\end{figure}

Fig.~\ref{fig:ImpulsiveWaveAlongRightBoundaryLambda0p10p2} shows the $\Psi_2$ and $\Psi_4$ components along the right boundary for $\lambda=0.6$ and $\lambda=1.2$, where a future singularity is formed when $\lambda=0.6$.

Like in the $\lambda=0$ case this is a curvature singularity. It is much easier to see this in the $\lambda > 0$ case as the Weyl invariant $I$ diverges to positive infinity.


\section{Two waves}
We now present results pertaining to the scattering of two colinearly polarized gravitational waves. The setup is analogous to the single wave case of Sec.~\ref{sec:SingleWave} with the exception of the boundary condition for $\Psi_4$, which is now taken to be $\Psi_4(t,-1) = p(u(t))$. We continue to use the gauge $F=\rho'-\rho$ which corresponds to the gauge used in the Khan-Penrose solution for colliding colinearly polarized impulsive gravitational plane waves with $\lambda=0$ \cite{khan1971scattering}. It is found that many features are similar to the case of one wave.

\subsection{Comparison against $\lambda=0$}
\begin{figure}[H]
    \centering
    \subfloat[\centering $\lambda=0$]
    {{\includegraphics[width=0.5\linewidth]{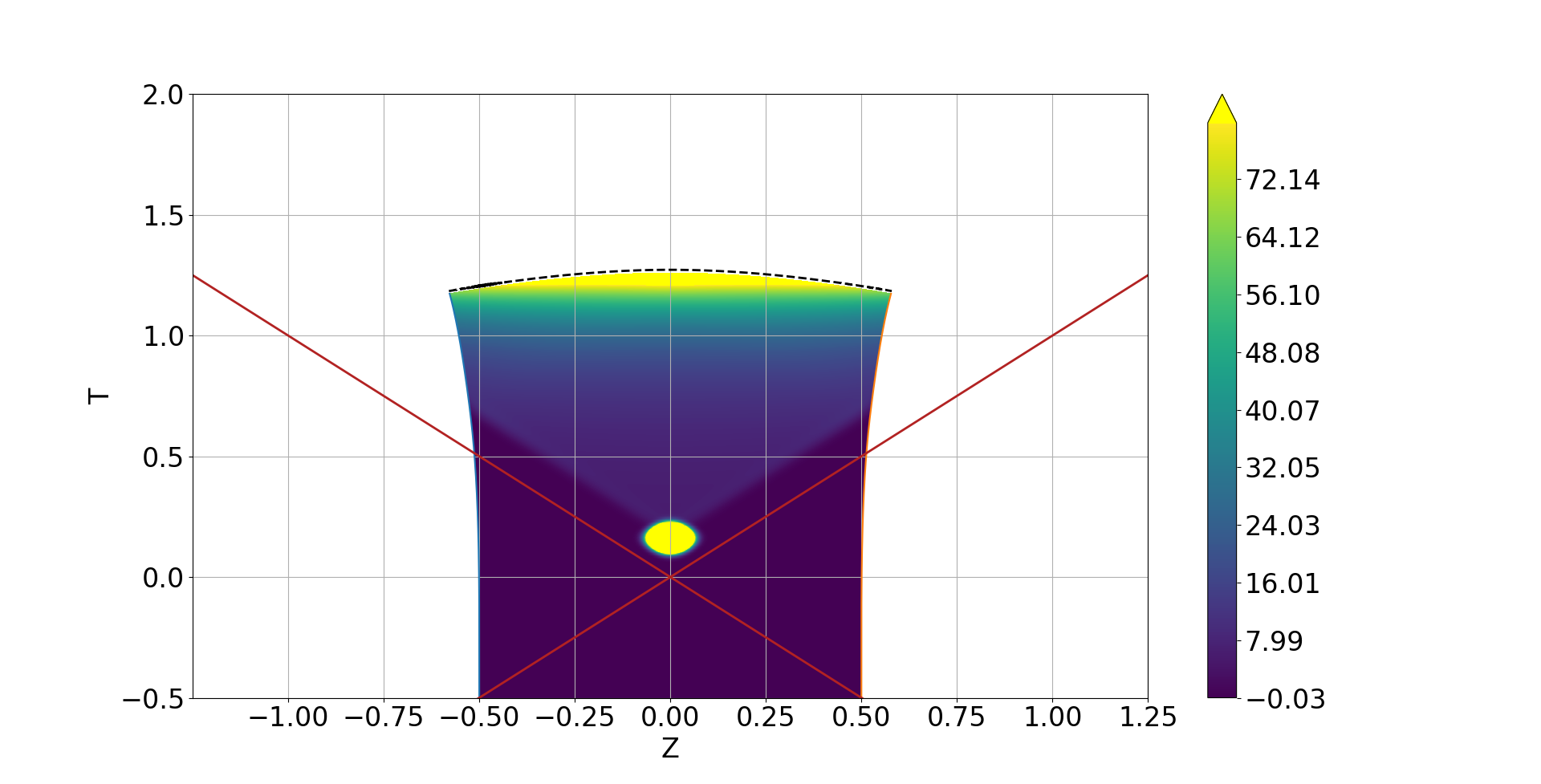}}}
    \qquad
    \subfloat[\centering $\lambda=0.06$]
    {{\includegraphics[width=0.5\linewidth]{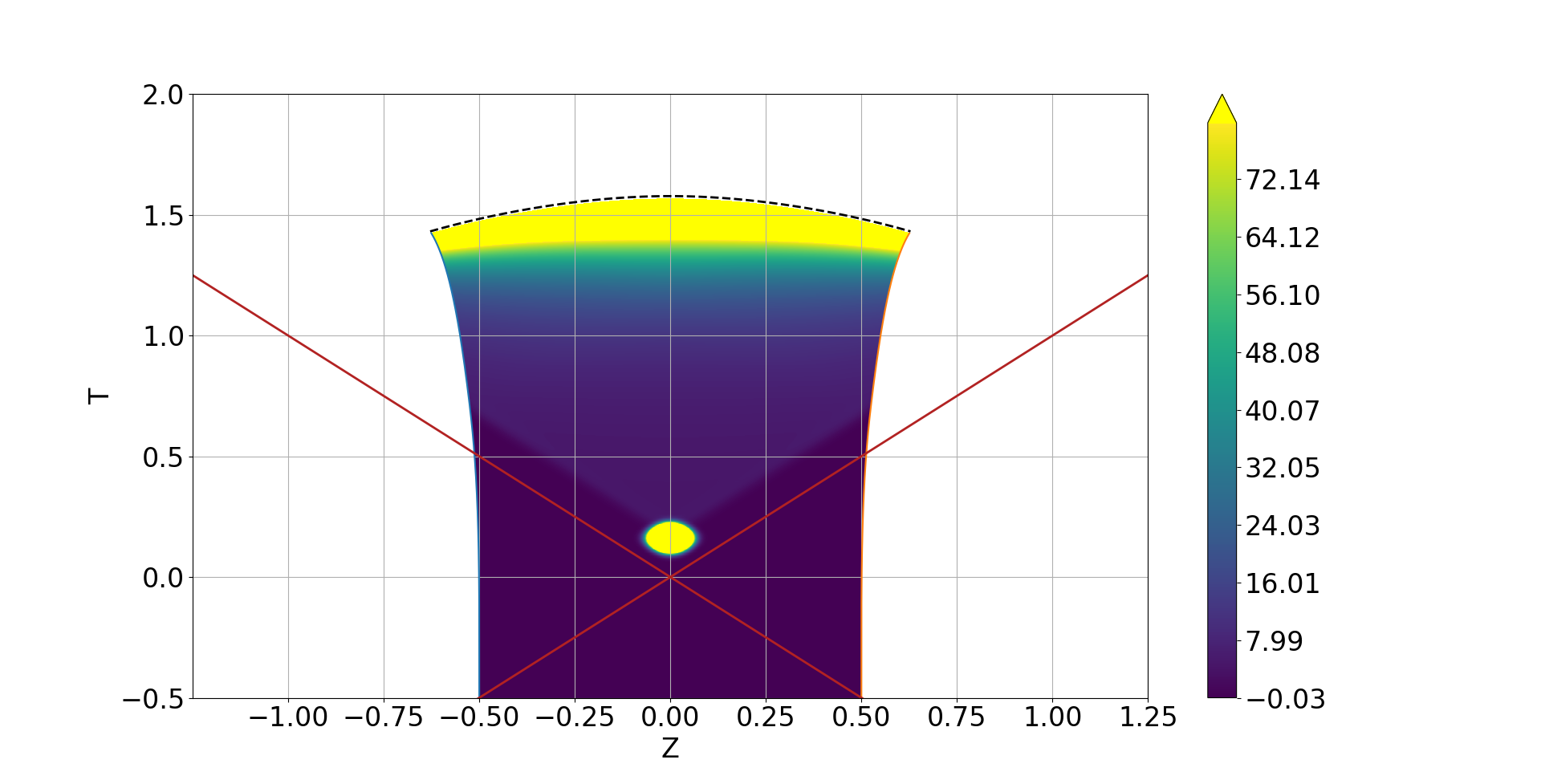}}}
    \\
    \subfloat[\centering $\lambda=0.6$]
    {{\includegraphics[width=0.5\linewidth]{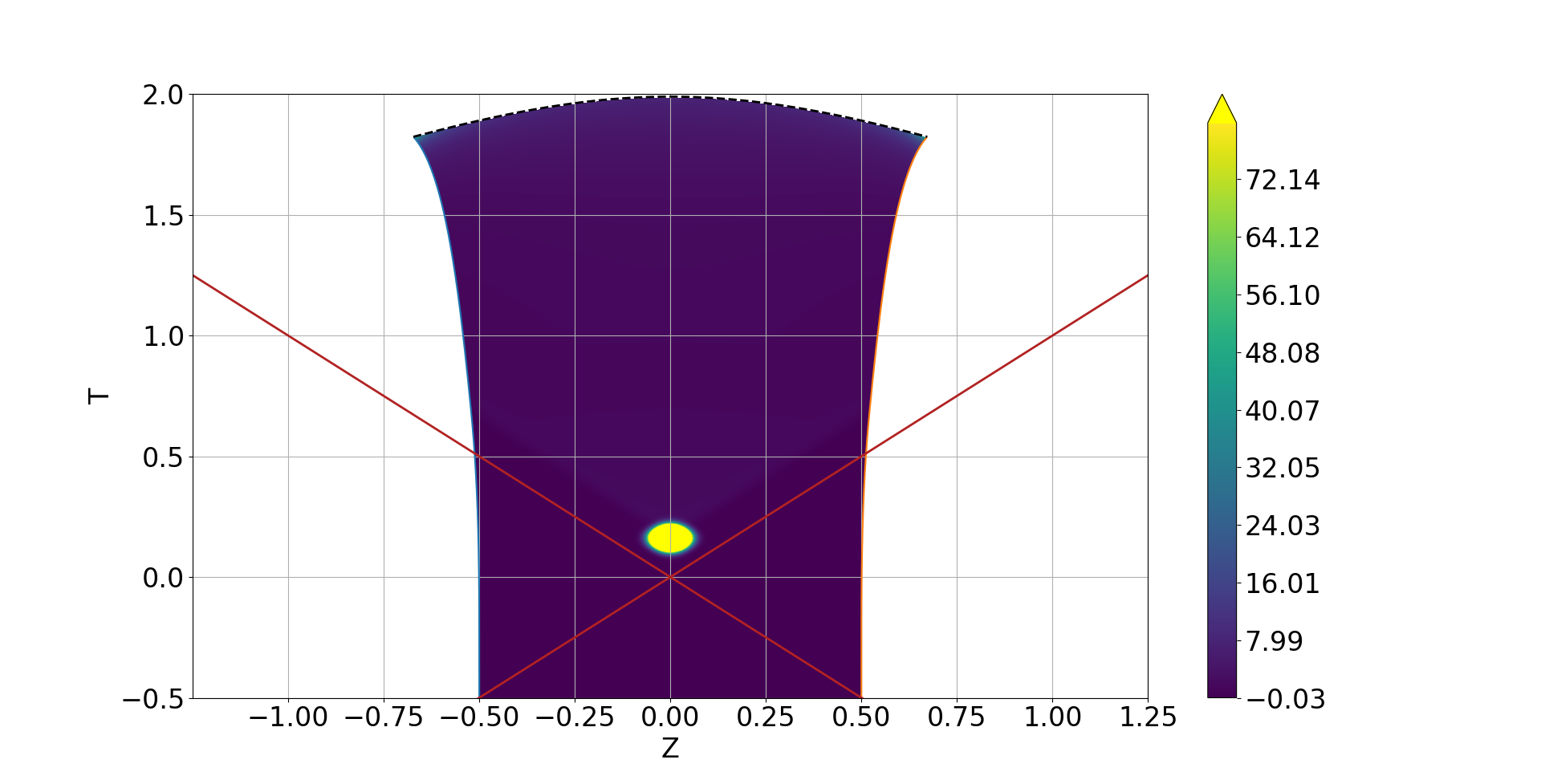}}}
    \qquad
    \subfloat[\centering $\lambda=6$]
    {{\includegraphics[width=0.5\linewidth]{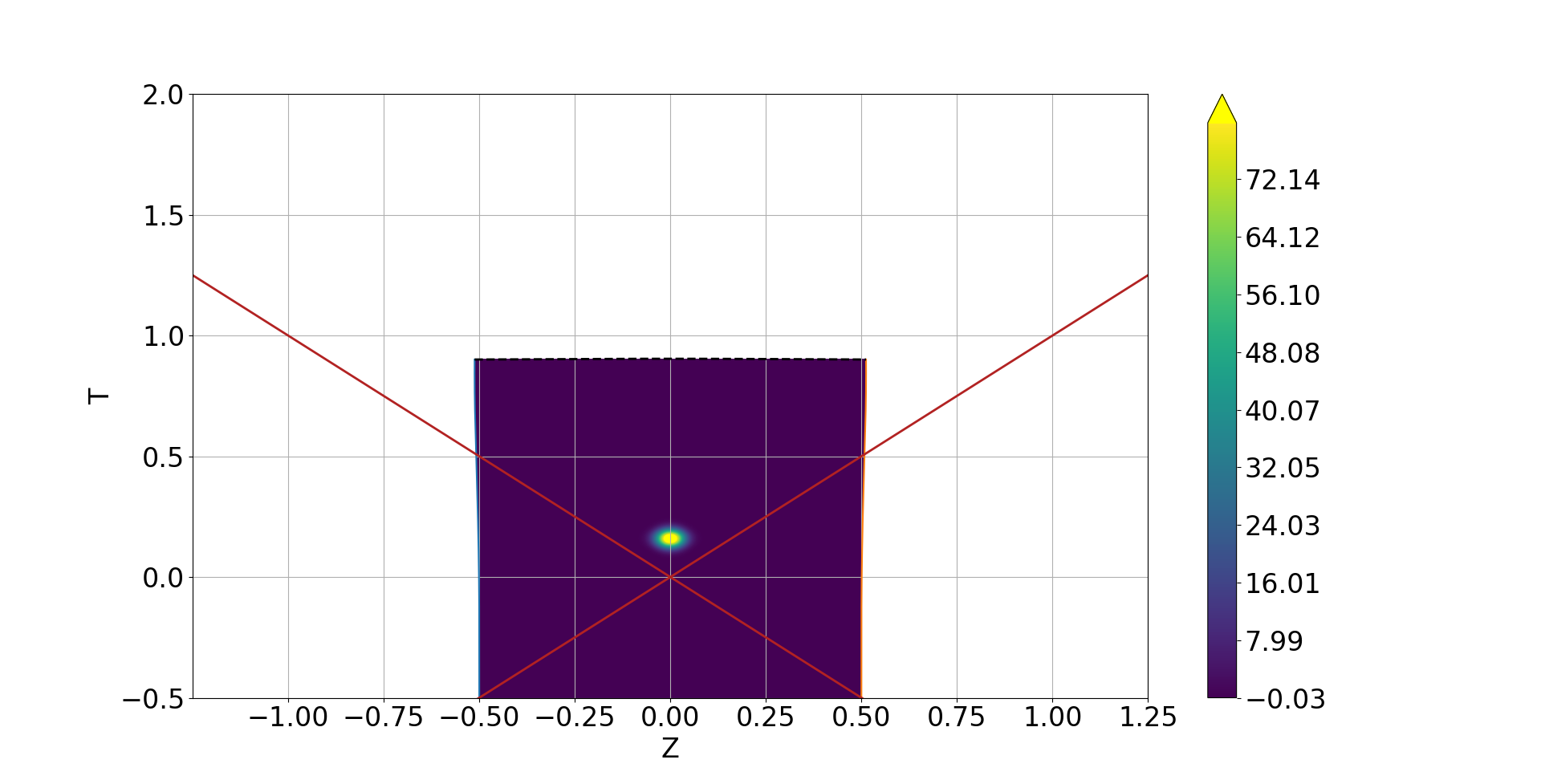}}}
    \caption{Penrose-Carter contour plots of the Weyl invariant $I$ for the case of colliding waves with varying $\lambda$.}
    \label{fig:CollidingWavesI}
\end{figure}

The general behaviour can be explained by looking at contour plots of $I$ in Fig.~\ref{fig:CollidingWavesI} for varying $\lambda$ (so that we can see how $\lambda>0$ differs from $\lambda=0$) and fixing $a=1$ in the wave profiles. If $\lambda$ is small enough ($\lambda=0$ or $\lambda=0.06$), we obtain a future curvature singularity. As $\lambda$ gets larger ($\lambda=0.6$), the expansion increases the time before this singularity occurs. If we increase $\lambda$ more ($\lambda=6$), we get to the situation where the expansion has wiped out the waves and the effect of their scattering on the curvature, and we asymptote back to dS again. 

\begin{figure}[H]
    \centering
    \subfloat[\centering $\lambda=0.06$]
    {{\includegraphics[width=0.5\linewidth]{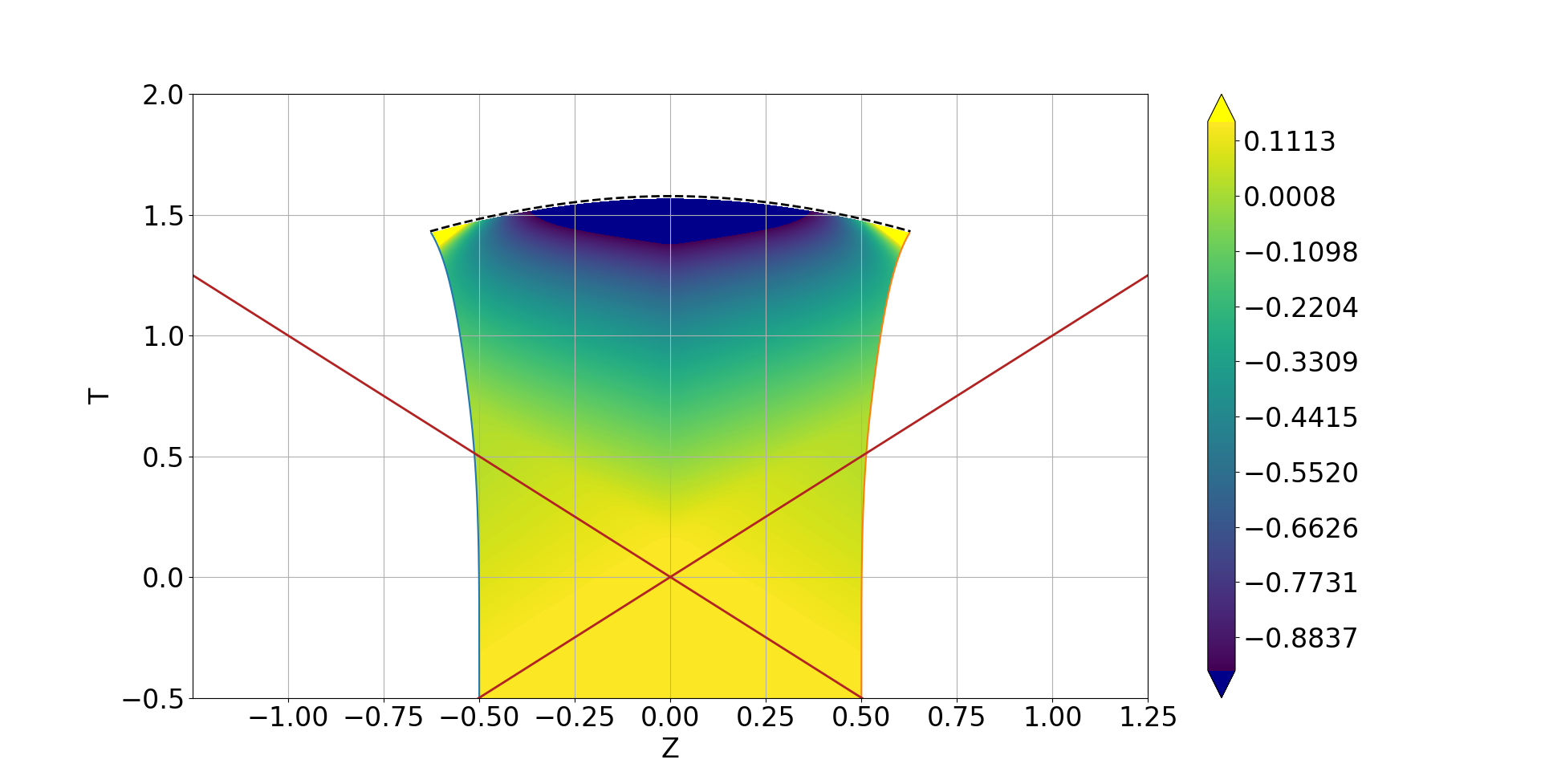}}}
    \qquad
    \subfloat[\centering $\lambda=6$]
    {{\includegraphics[width=0.5\linewidth]{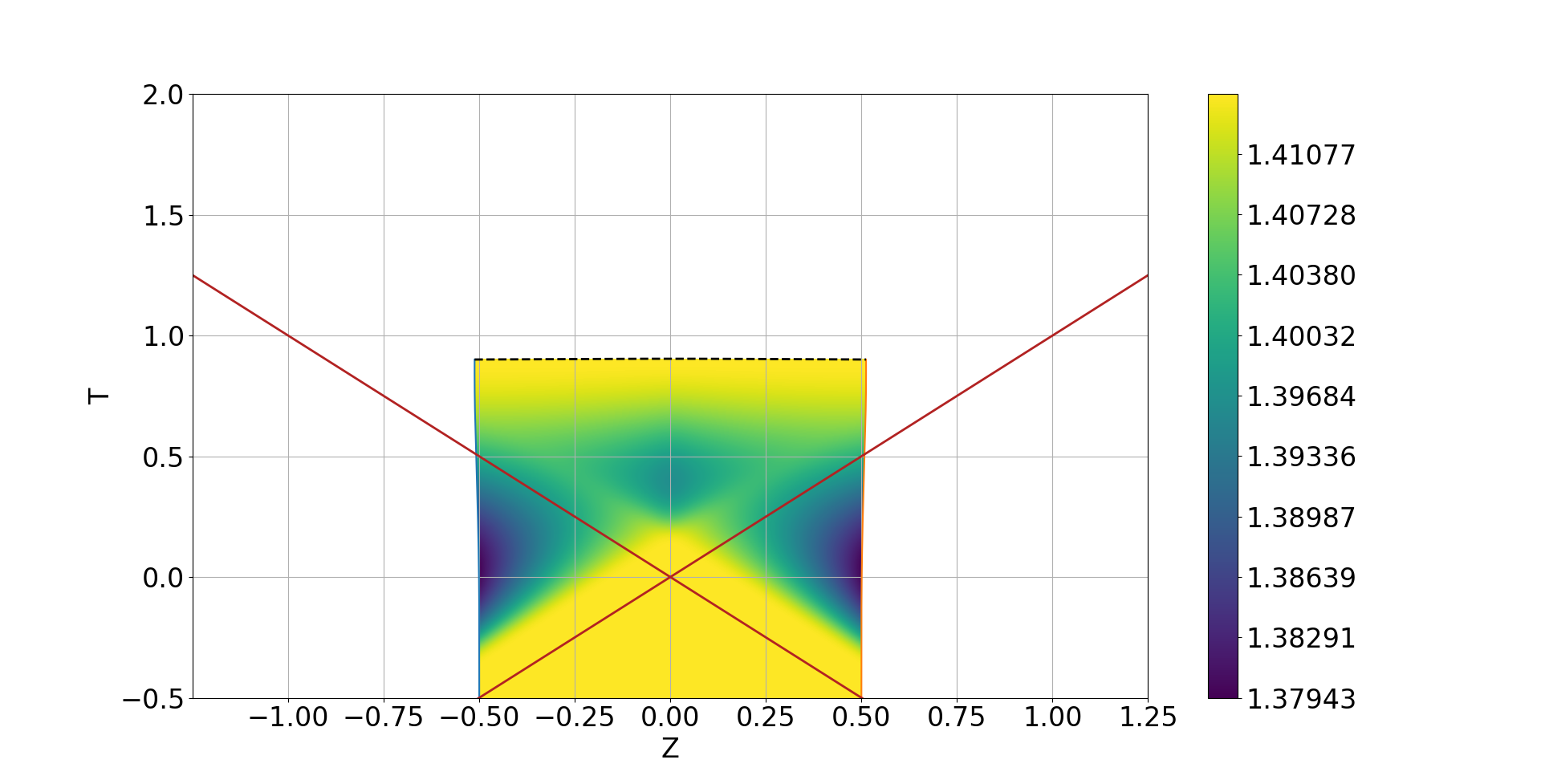}}}
    \caption{Penrose-Carter contour plots of the expansion rate $\mathcal{H}$ for the case of colliding waves with varying $\lambda$.}
    \label{fig:CollidingWavesH}
\end{figure}

Fig.~\ref{fig:CollidingWavesH} shows the expansion rate $\mathcal{H}$ decreasing the most in the centre of the collision, $u=v$, where the Weyl invariant $I$ attains a local maximum (in time and space.)

\subsection{Critical behaviour}\label{sec:criticalbehaviourtwowaves}
\begin{figure}[H]
    \centering
    \includegraphics[width=0.3\linewidth]{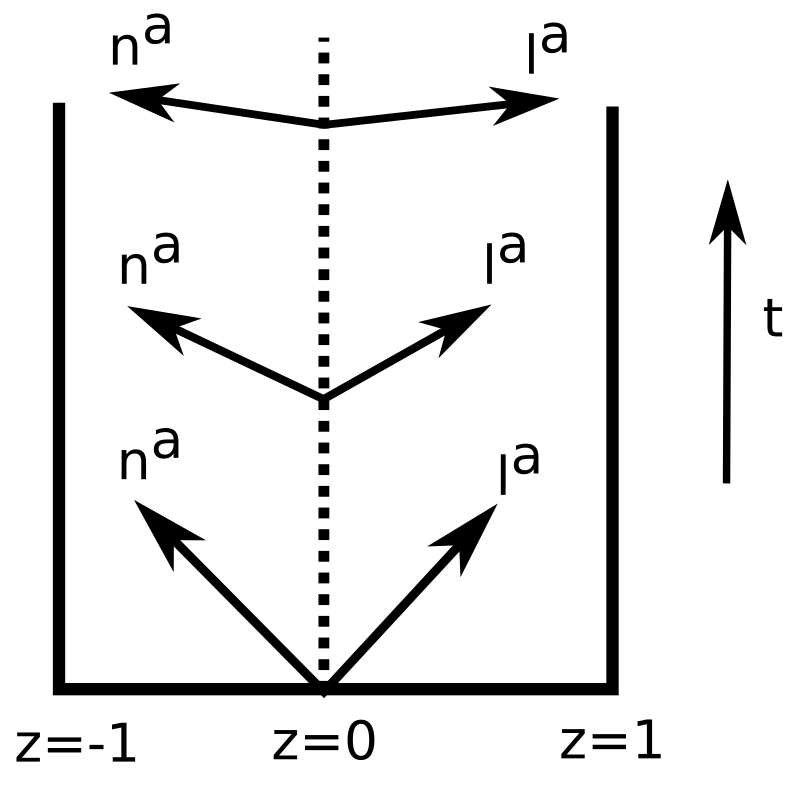}
    \caption{The effect of $A\rightarrow\infty$ as $t\rightarrow\infty$ on the null vectors along a $z=\;$constant curve.}
    \label{fig:AToInfinityDiagram}
\end{figure}

Now we fix $\lambda=0.6$ and see how varying the area of the wave profile $a$ affects things. We find the following three scenarios, where $a_1$ and $a_2$ are given later in the section, and are found with binary search:
\begin{itemize}
    \item[1] $a\lessapprox a_1$: asymptote back to dS.
    \item[2] $a_1\lessapprox a \lessapprox a_2$: $\mu\rightarrow\infty$ but $I\rightarrow0$.   
    \item[3] $a\gtrapprox a_2$: $\mu\rightarrow\infty$ and $I\rightarrow\infty$.
\end{itemize}
Only in case 3 do $\rho,\,\rho',\,\sigma,\,\sigma'$ diverge, in the other two they asymptote back to their initial values. Due to the evolution equation $\sqrt{2}\del_tA = (\mu + \bar{\mu})A$, in cases 2 and 3 we have that $A\rightarrow\infty$ also, causing the $t,z$ portion of the line element to approach $\textrm{d}t^2$, causing an infinite contraction in the $z$-direction. This is represented in $l^a$ and $n^a$ as shown in Fig.~\ref{fig:AToInfinityDiagram}, where the $t=\;$constant surfaces approach being null. Further, as we discovered in Sec.~\ref{sec:general-setup}, the real part of $\mu$ is essentially the acceleration of the unit conormal to the $z=\;$constant surfaces and the fact that this acceleration diverges to negative infinity agrees with the contraction in this direction.

We are in the Gau\ss\; gauge, and along spatially constant curves, which are in this case geodesics, the proper time and the time $t$ are equivalent. Our gauge can then be thought of as adapted to free falling observers. This then lends the physical interpretation of the caustic singularity in case 2. The three possible futures occurring after the interaction of the gravitational waves with these observers can then be described as follows:
\begin{itemize}
    \item Case 1: The gravitational contraction is not strong enough to cause the timelike geodesics to converge or the curvature to diverge.
    \item Case 2: The gravitational contraction is strong enough to cause the timelike geodesics to converge and create a coordinate singularity. However, it is not strong enough to cause the curvature to diverge and this goes back to zero.
    \item Case 3: The gravitational contraction is strong enough to cause both the timelike geodesics to converge and the curvature to diverge, resulting in a physical curvature singularity.
\end{itemize}

It is noted that in the gauge $B=0$ the characteristic speeds of the waves are $\pm A$. In the cases where $A\rightarrow\infty$ we decrease the CFL number $c$ dynamically to avoid instabilities and settle with smaller timesteps instead.

As we now have \emph{two} bifurcations, which we call $a_1$ and $a_2$, it remains to be seen whether these will have critical behaviours.

\begin{figure}[H]
    \centering
    \subfloat[\centering $\mathcal{H}$]
    {{\includegraphics[width=0.5\linewidth]{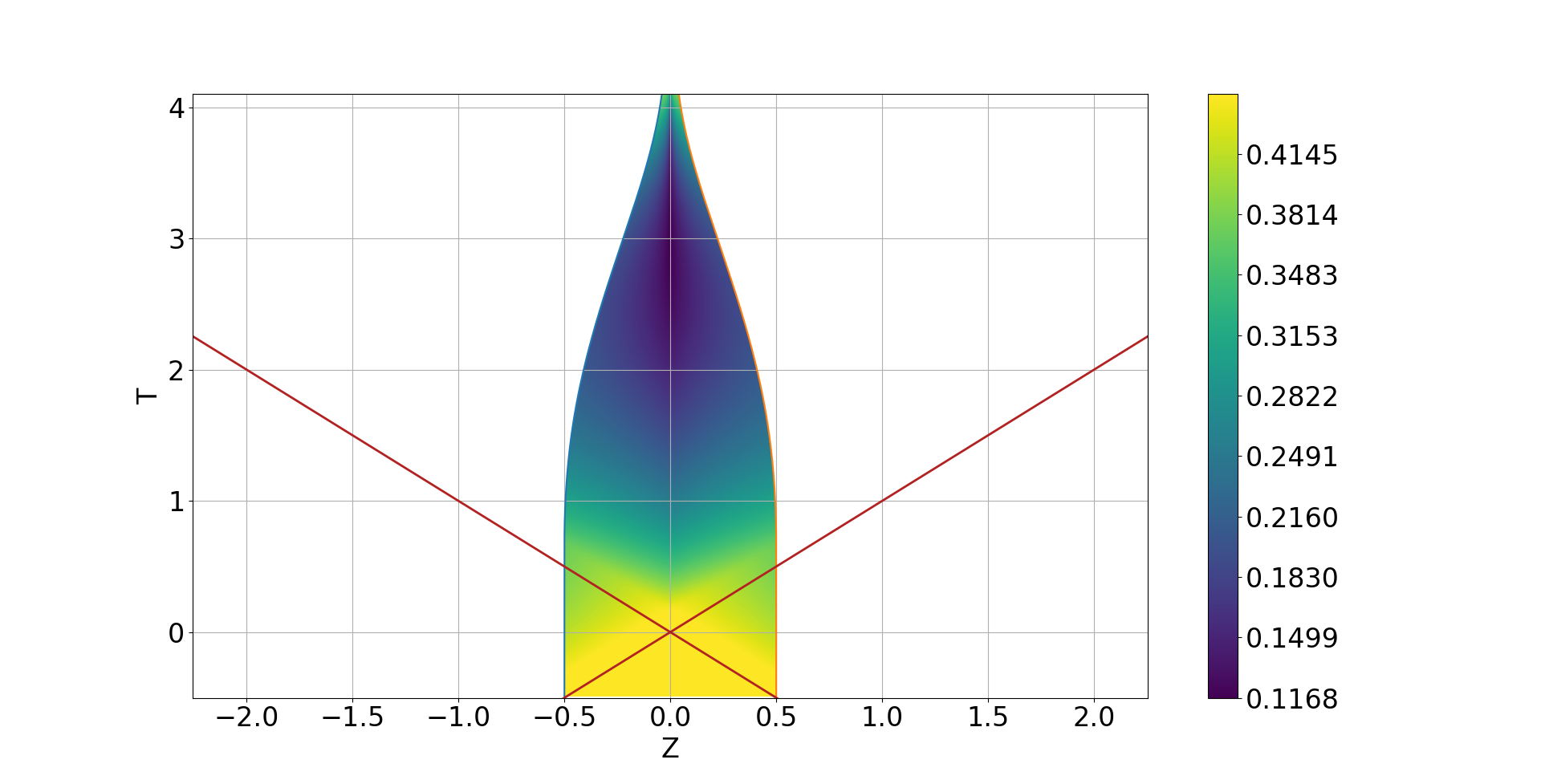}}}
    \qquad
    \subfloat[\centering $\mathcal{I}$]
    {{\includegraphics[width=0.5\linewidth]{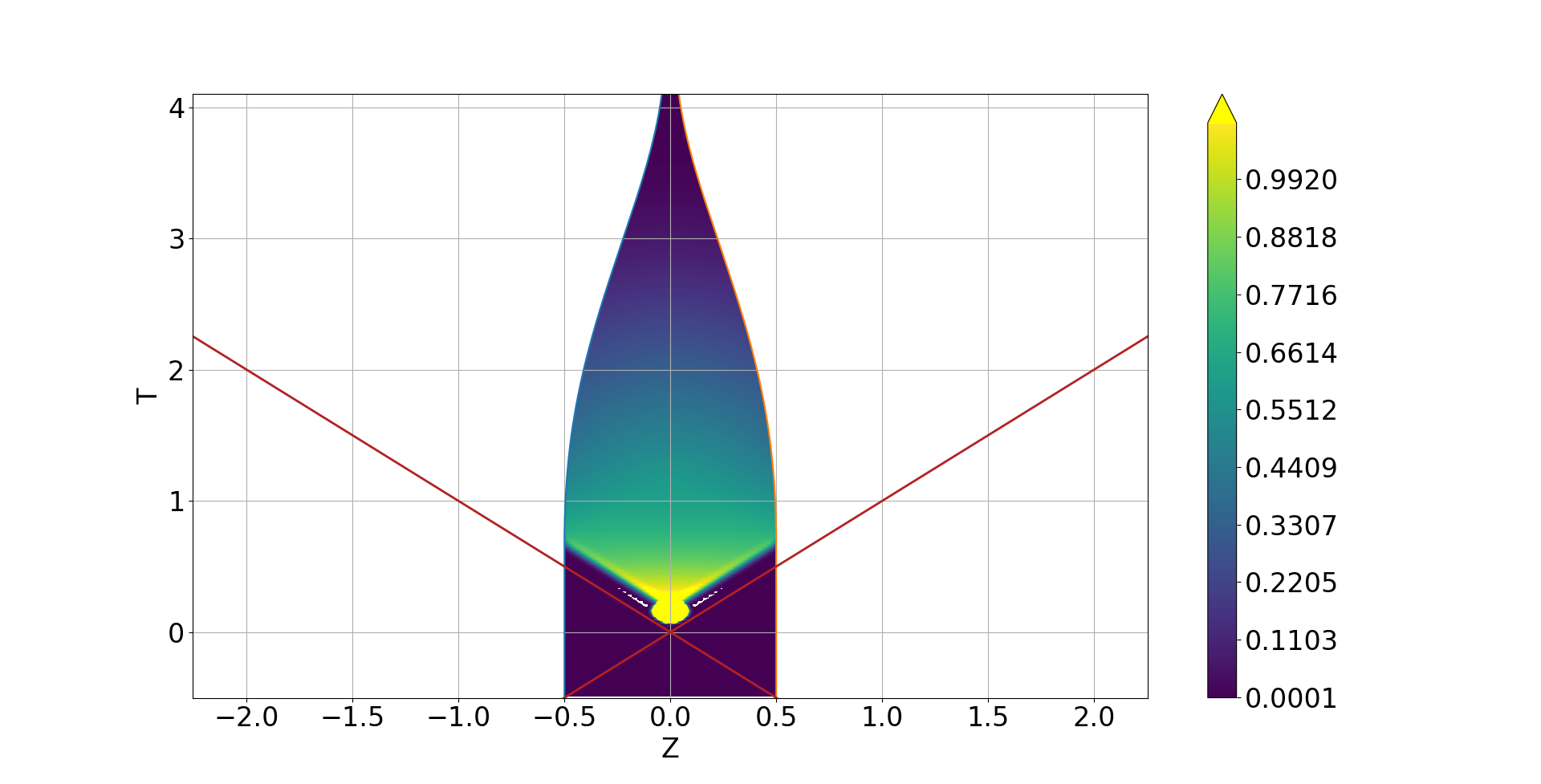}}}
    \caption{The expansion rate $\mathcal{H}$ and Weyl invariant $I$ with $a\approx a_1$ and $\lambda=0.6$.}
    \label{fig:HIMuCrit}
\end{figure}

\begin{figure}[H]
    \centering
    \subfloat[\centering $A$]
    {{\includegraphics[width=0.4\linewidth]{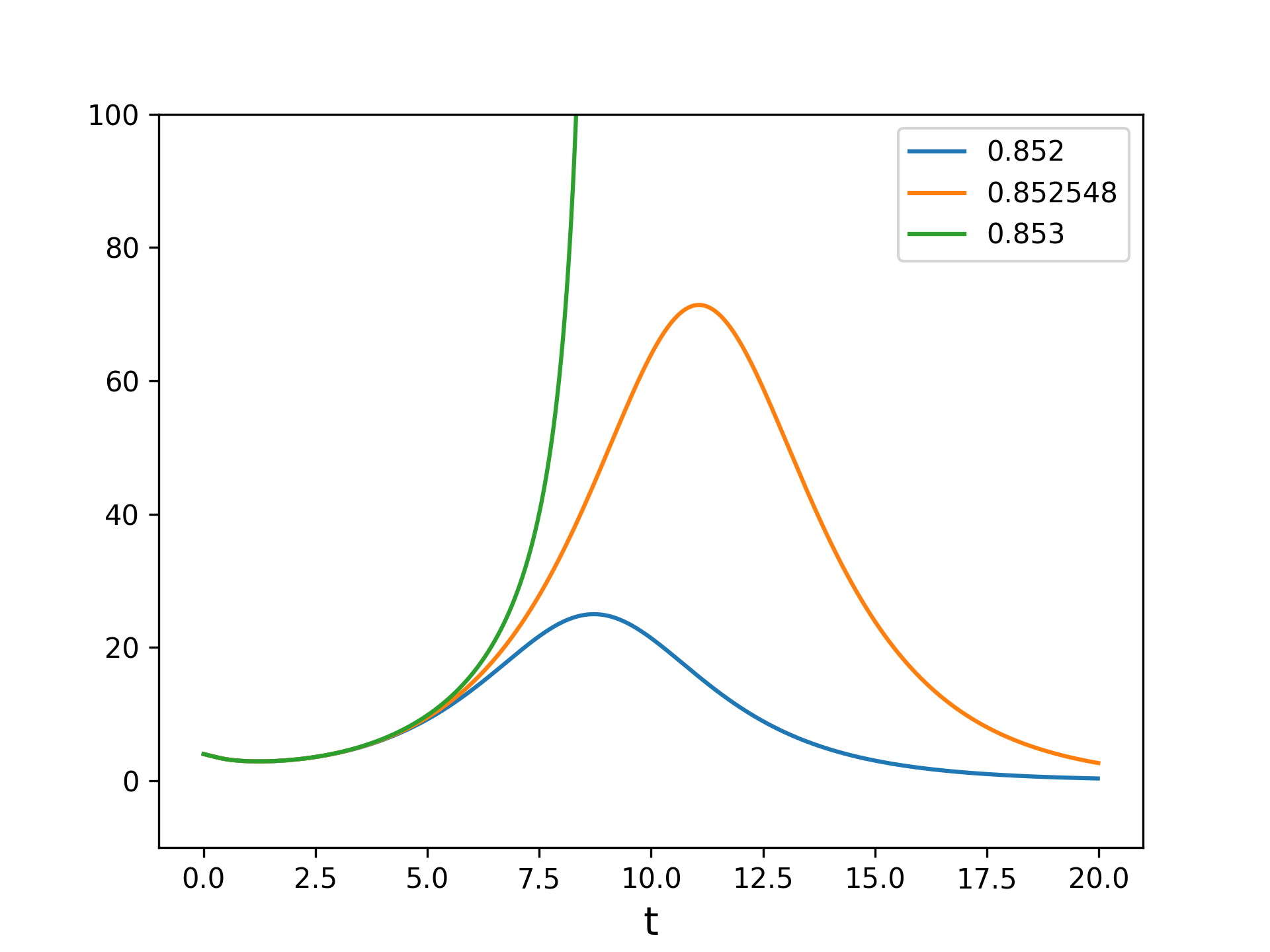}}}
    \qquad
    \subfloat[\centering $\mathcal{H}$]
    {{\includegraphics[width=0.4\linewidth]{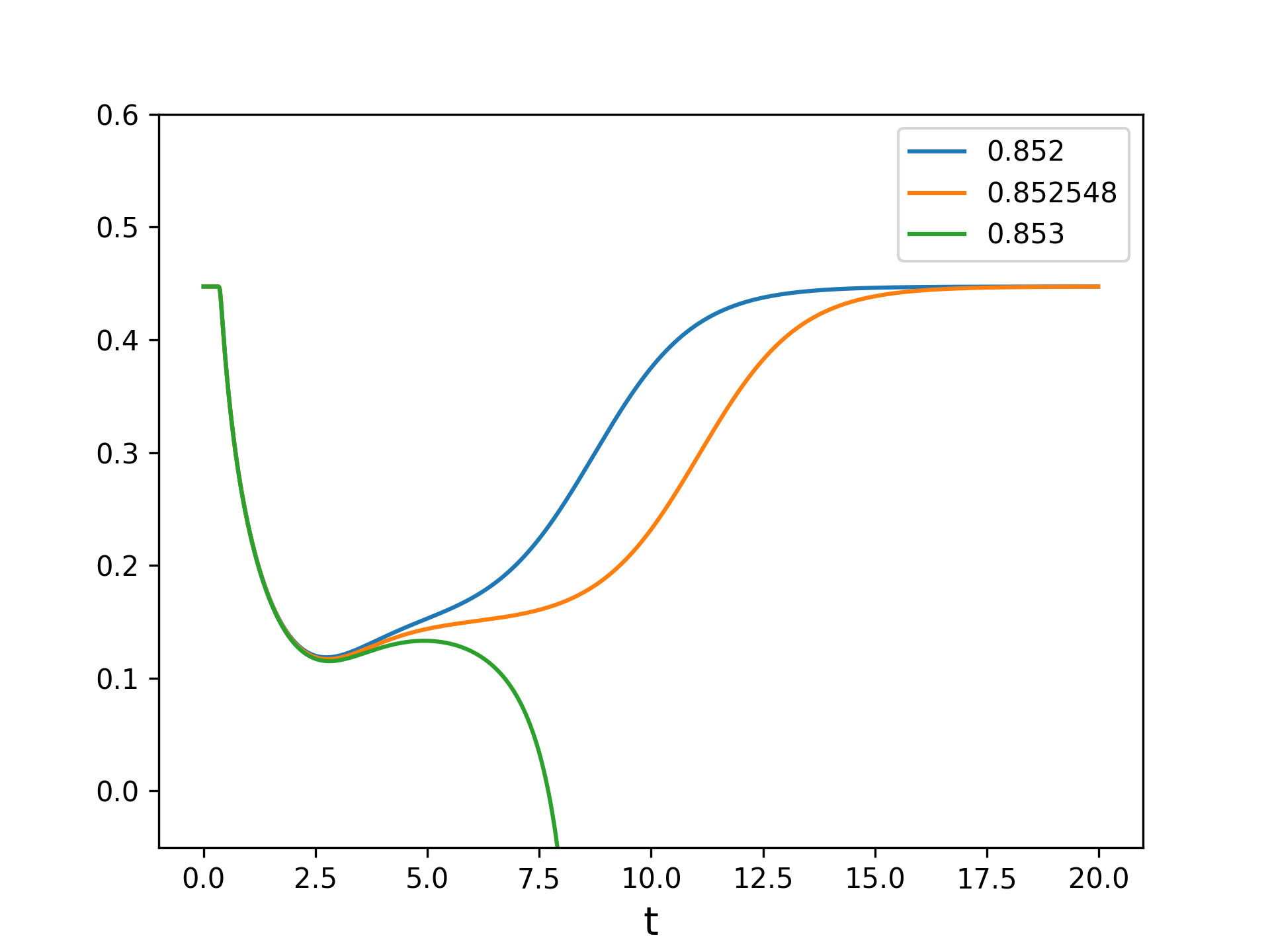}}}
    \caption{The metric function $A$ and the expansion rate $\mathcal{H}$ with $\lambda=0.6$ along $u=v$ equiv. $z=0$ for multiple values of $a$ close to $a_1$.}
    \label{fig:AHAlongz0MuCrit}
\end{figure}

We find, again using binary search, that approximately $a_1 \approx 0.852548$. Unlike in Sec.~\ref{sec:criticalbehaviour}, the constraints do not diverge as our simulations use a wave profile with area $a$ closer to $a_1$. Fig.~\ref{fig:HIMuCrit} and Fig.~\ref{fig:AHAlongz0MuCrit} show that the expansion rate drops to around $25\%$ of its original value at its minimum, for a long time, before asymptoting back to dS again. This implies that we can cause, with just two colliding waves, the expansion rate to locally decrease substantially for a certain period, without causing a future singularity. Further, it is noticed that although $\mu$ differs substantially in the above cases, $\rho, \rho', \sigma$ and $\sigma'$ change very little, and if drawn differ by an amount smaller than the drawn curve.

\begin{figure}[H]
    \centering
    \subfloat[]
    {{\includegraphics[width=0.4\linewidth]{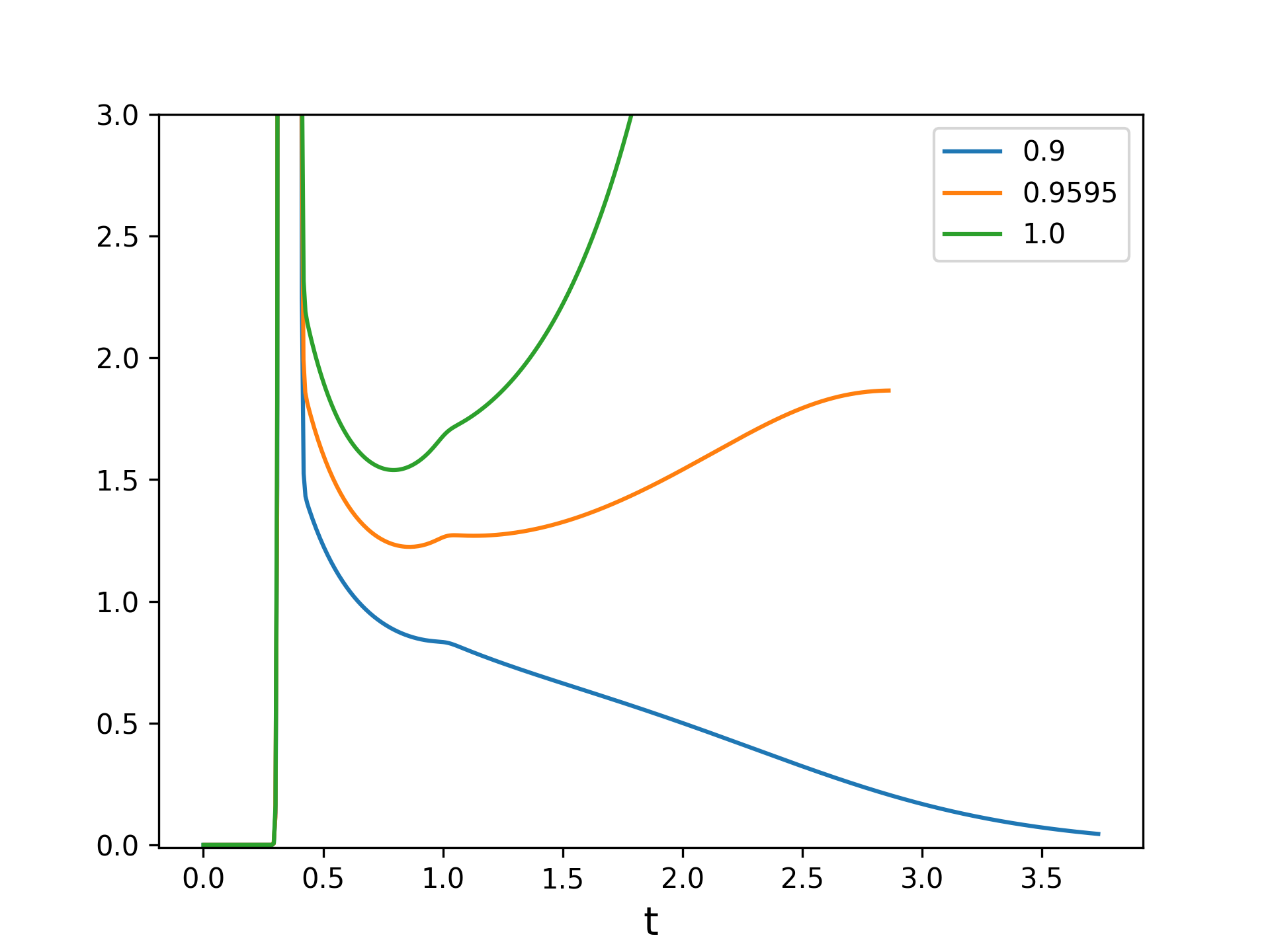}}}
    \qquad
    \subfloat[]
    {{\includegraphics[width=0.4\linewidth]{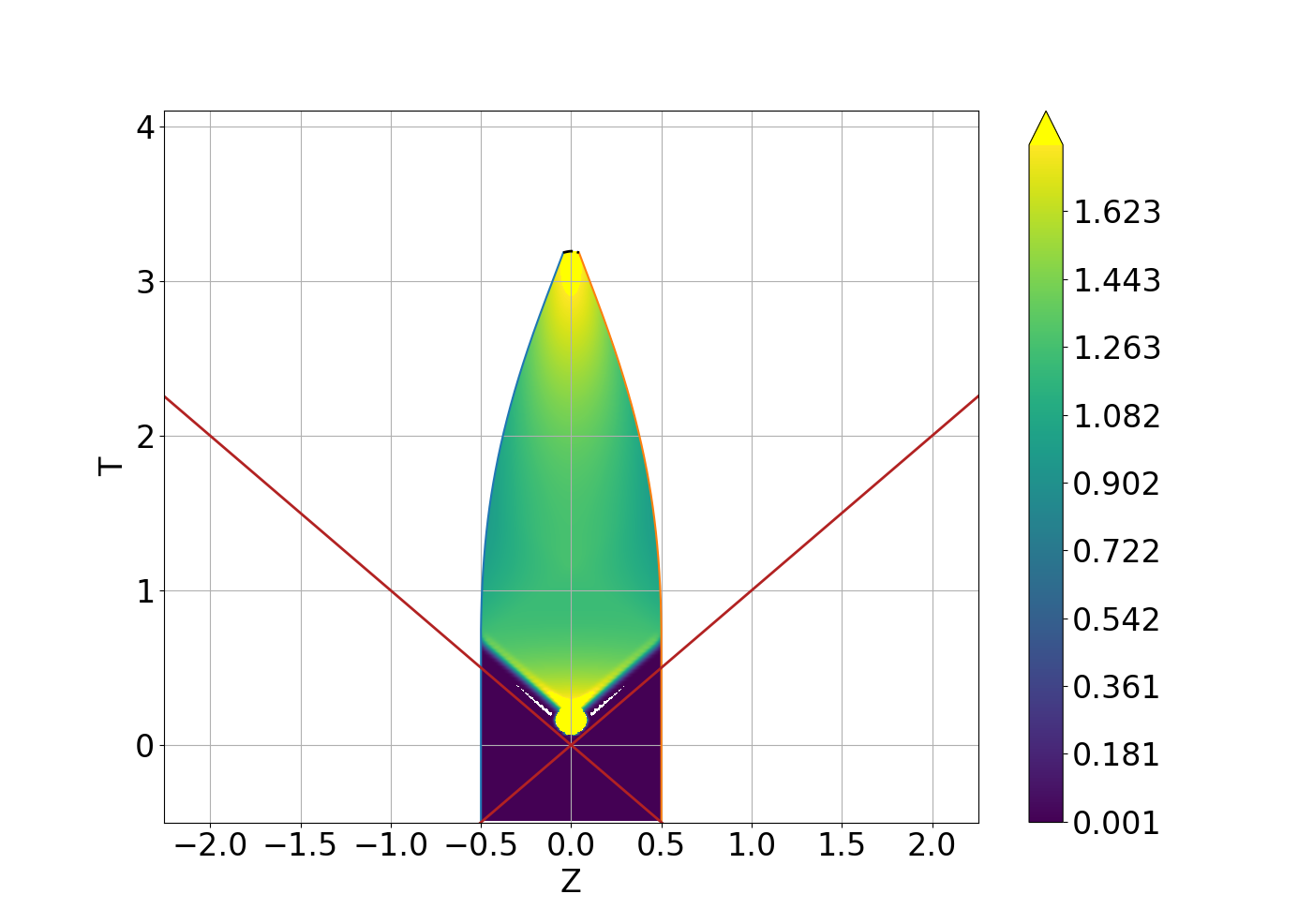}}}
    \caption{(a) The Weyl invariant $I$ with $\lambda=0.6$ along $u=v$, i.e. $z=0$, for multiple values of $a$ close to $a_2$ and (b) a contour plot of $I$ for $a=0.9595$.}
    \label{fig:IAlongz0ICrit}
\end{figure}

We find, again using binary search, that approximately $a_2\approx0.9595$, and find that taking $a$ close to this value results in the constraints remaining well behaved. Fig.~\ref{fig:IAlongz0ICrit} shows that the Weyl invariant $I$ diverges for $a>a_2$, goes to zero for $a<a_2$ and goes to some other value when $a\approx a_2$. In all these cases $\mu$ diverges to infinity and thus so does $A$. This implies that to maintain a stable evolution our timestep must decrease to compensate, and the simulations shown in Fig.~\ref{fig:IAlongz0ICrit} stop when the timestep becomes smaller than 1e-8. It is likely that the simulation with $a=0.9595$ does not converge to some constant value other than zero, but rather we cannot march in time far enough to see it either diverge to infinity or approach zero.

\subsection{Impulsive waves}\label{sec:CollidingImpulsiveWaves}
As in Sec.~\ref{sec:OneImpulsiveWave} we mimic the Dirac delta function wave profiles of the $\lambda=0$ solutions. For colliding waves, this is when $\Psi_0 = \delta(v)$ and $\Psi_4 = \delta(u)$. We thus choose our wave profiles as $\Psi_0(v,1) = q(v)$, $\Psi_4(u,-1) = q(u)$, where $q(x)$ is given in Eq.~\eref{eq:qBC} and approximates the Dirac delta function. Our results in Sec.~\ref{sec:criticalbehaviourtwowaves} indicate that we should explore three possible regions, namely regions where we asymptote back to dS, $\mu$ diverges but not $I$, and where $I$ diverges. These still exist for the approximately impulsive wave profiles and are exemplified by choosing $\lambda=6,\,0.72$ and $0.6$ respectively.

\begin{figure}[H]
    \centering
    \subfloat[\centering $\Psi_2$ for $\lambda=0.06$]
    {{\includegraphics[width=0.5\linewidth]{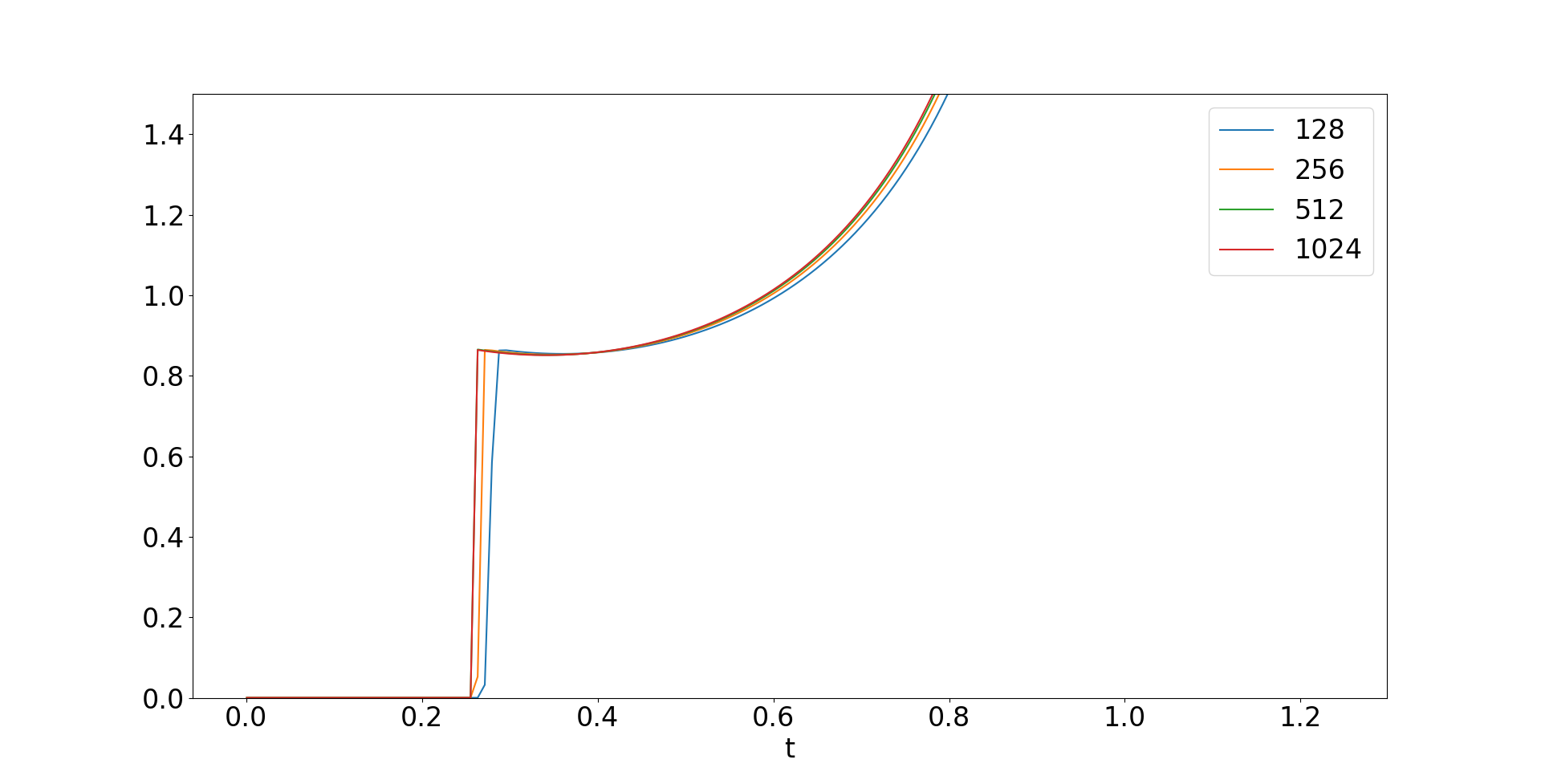}}}
    \qquad
    \subfloat[\centering $\Psi_4$ for $\lambda=0.06$]
    {{\includegraphics[width=0.5\linewidth]{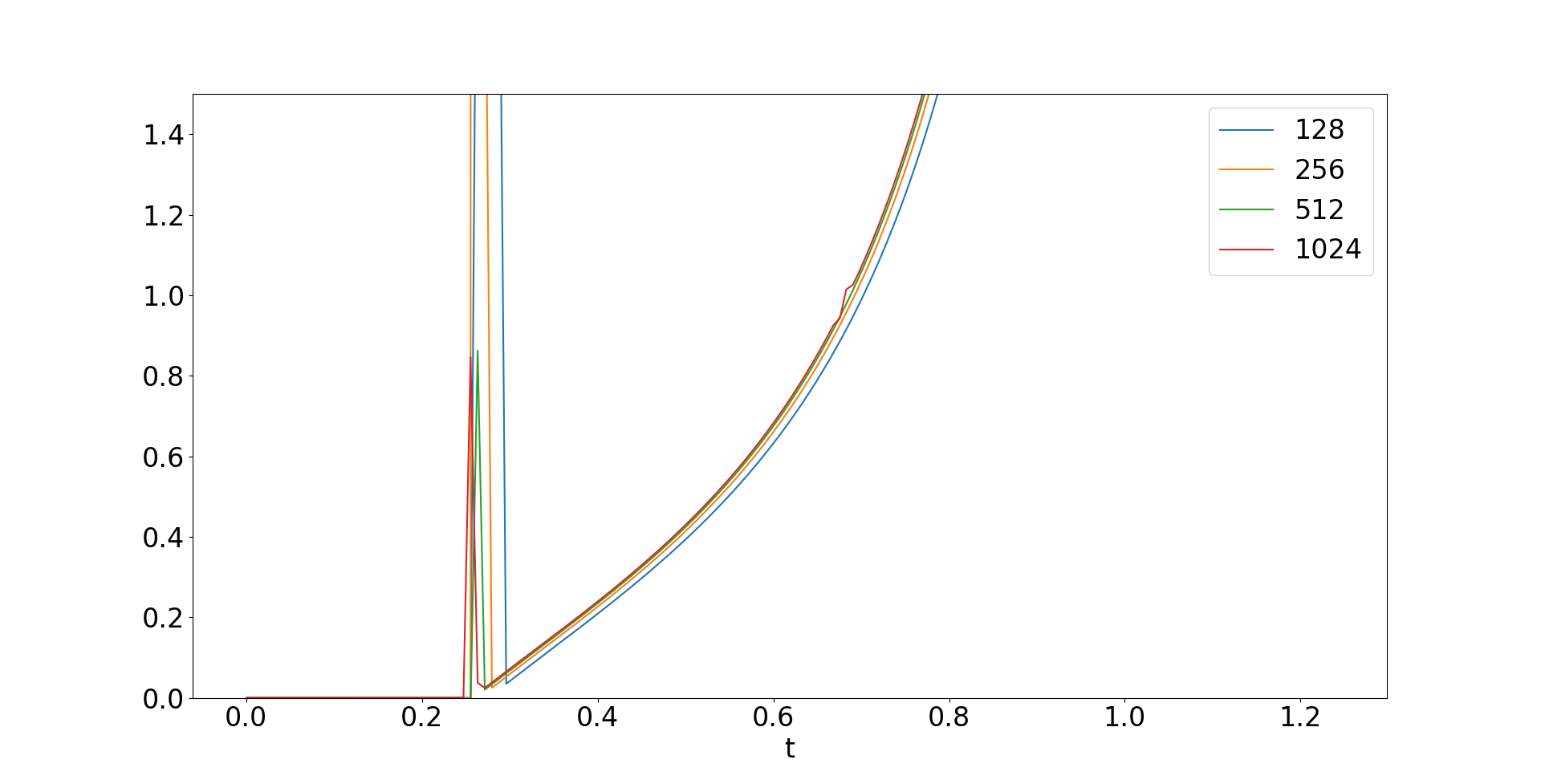}}}
    \\
    \subfloat[\centering $\Psi_2$ for $\lambda=0.72$]
    {{\includegraphics[width=0.5\linewidth]{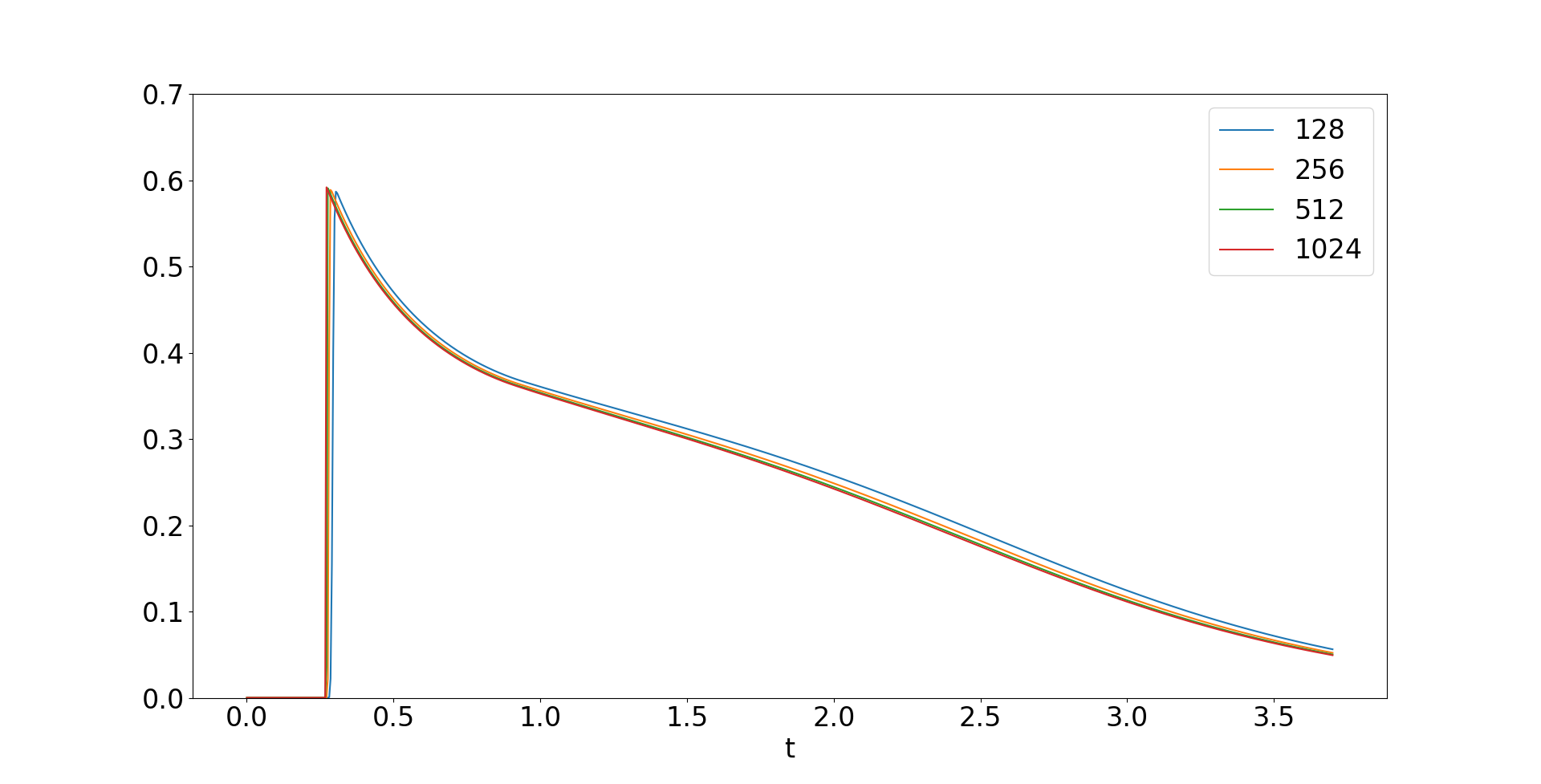}}}
    \qquad
    \subfloat[\centering $\Psi_4$ for $\lambda=0.72$]
    {{\includegraphics[width=0.5\linewidth]{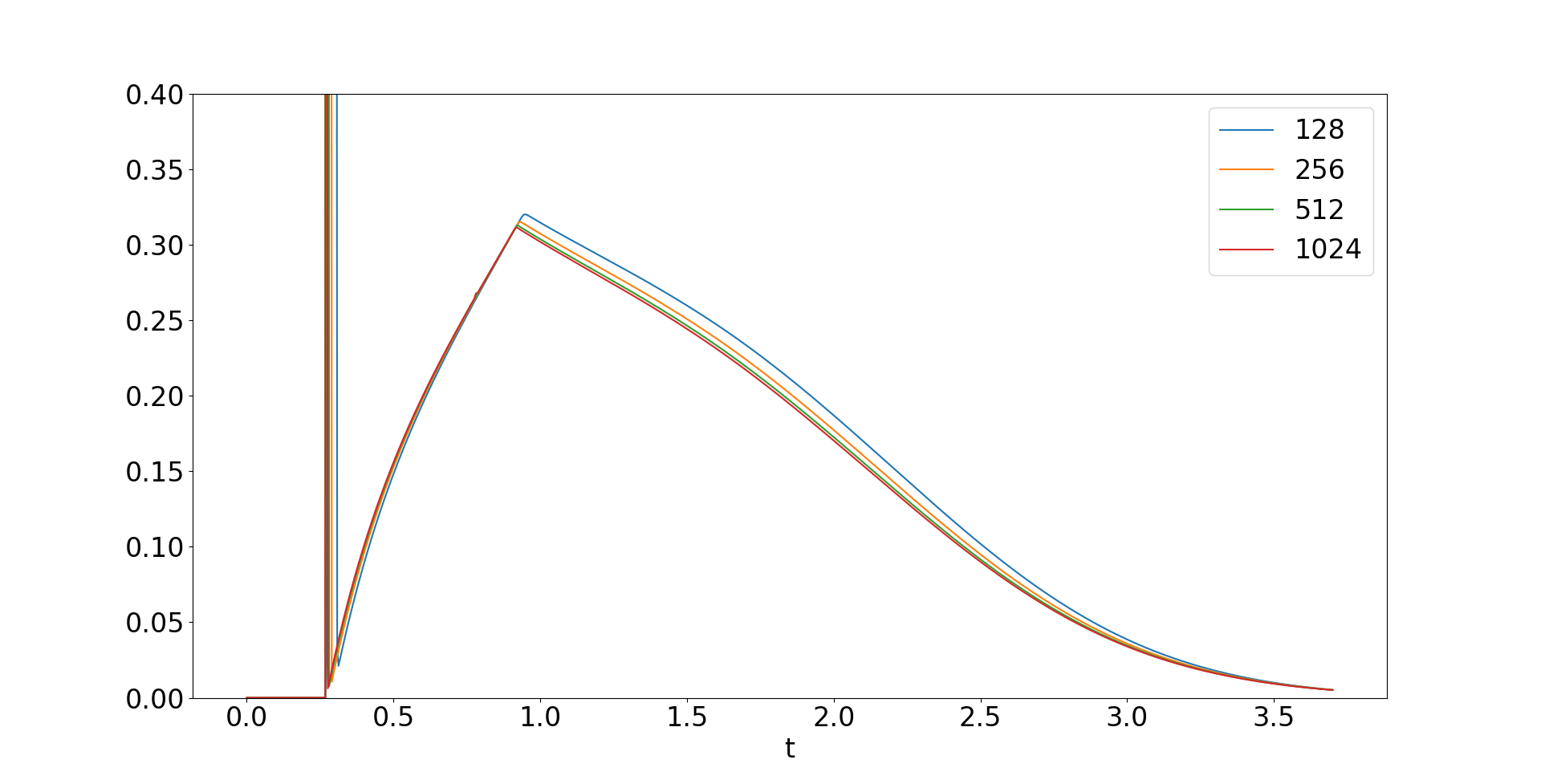}}}
    \\
    \subfloat[\centering $\Psi_2$ for $\lambda=6$]
    {{\includegraphics[width=0.5\linewidth]{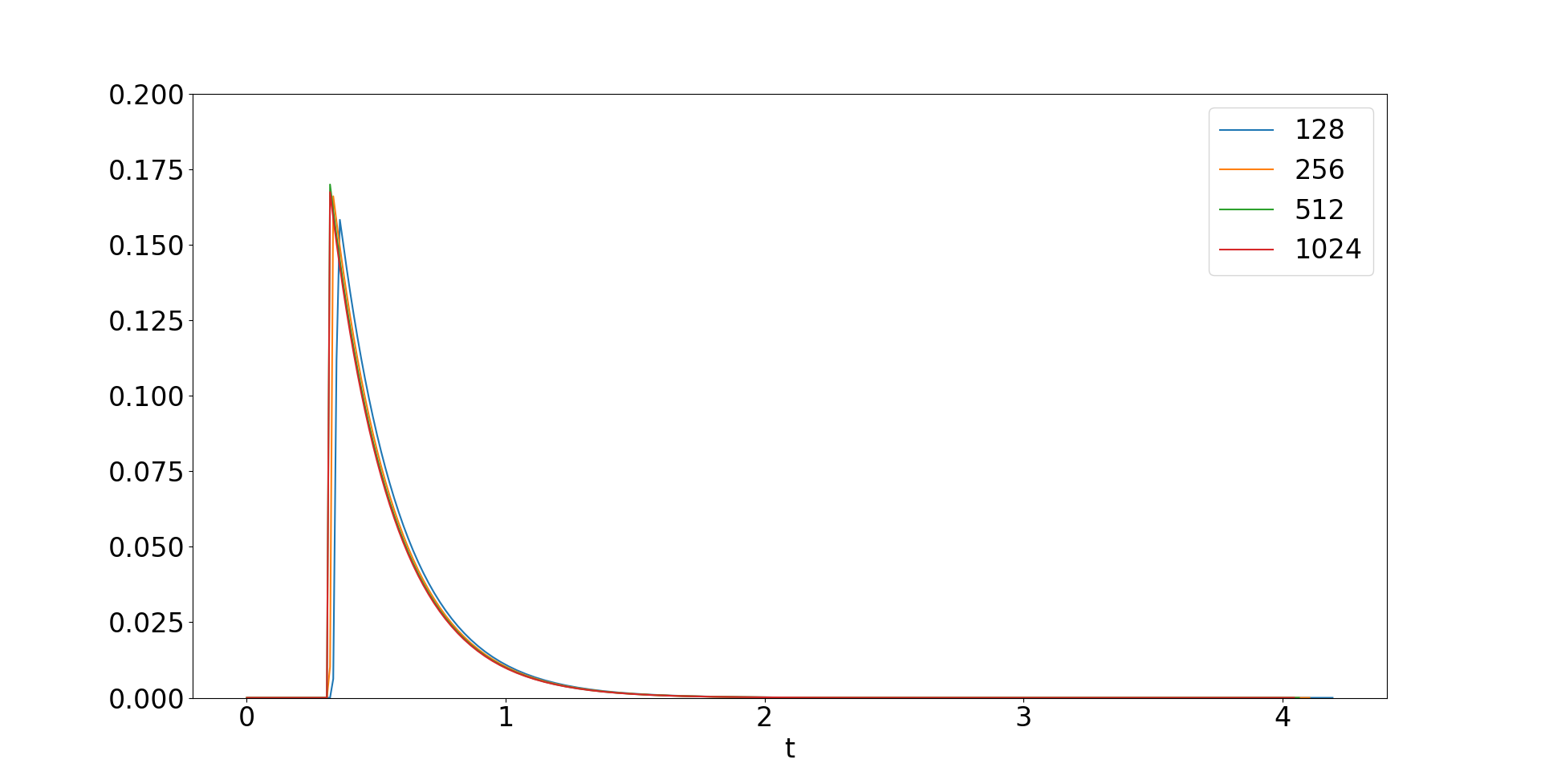}}}
    \qquad
    \subfloat[\centering $\Psi_4$ for $\lambda=6$]
    {{\includegraphics[width=0.5\linewidth]{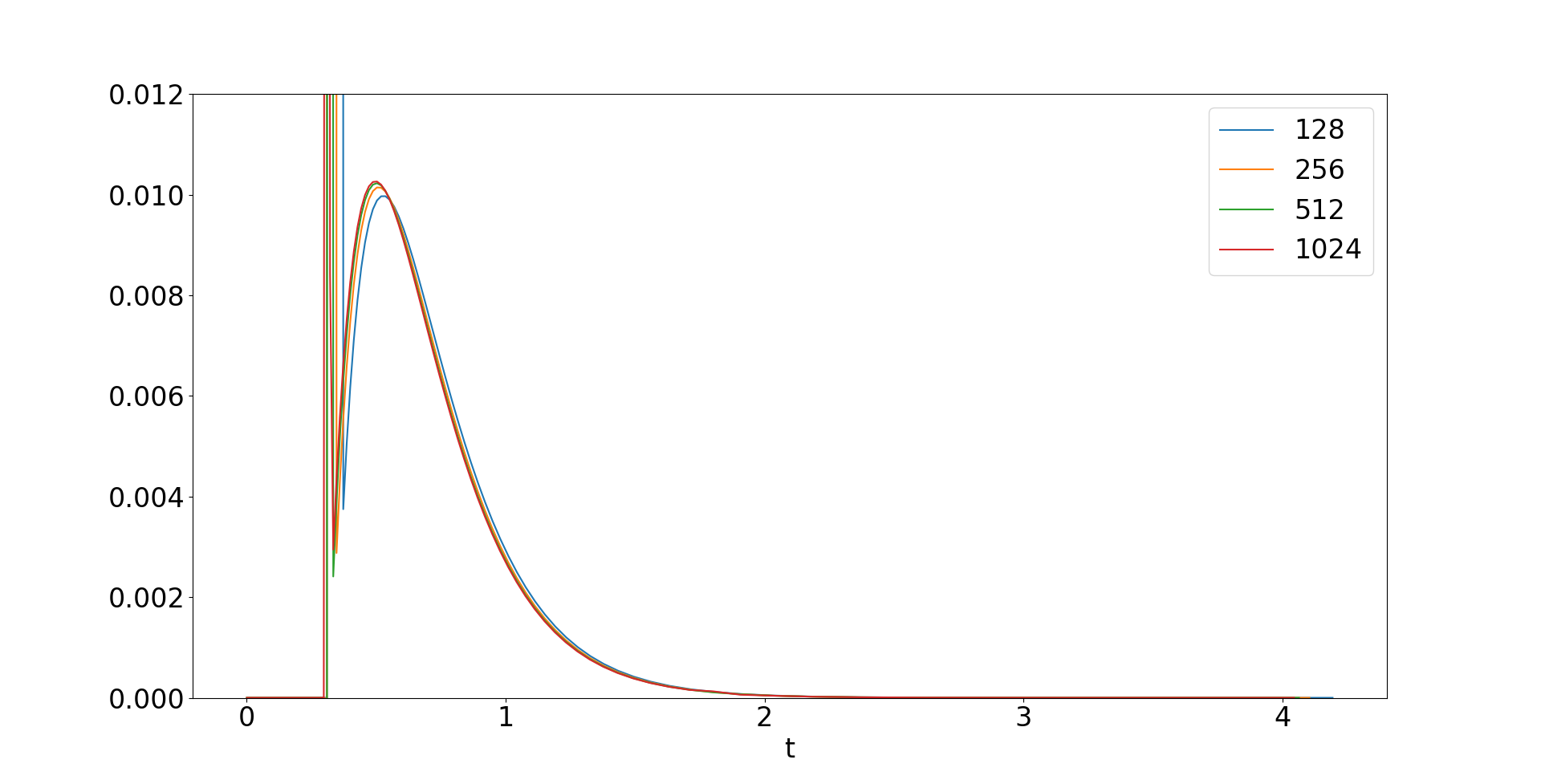}}}
    \caption{Plots along $u=v$ equiv. $z=0$ of $\Psi_2$ and $\Psi_4$ as $a\rightarrow\infty$ with $\lambda=0.6,\,0.72$ and $6$.}
    \label{fig:CollidingImpulsiveWavesAlongz0}
\end{figure}

Fig.~\ref{fig:CollidingImpulsiveWavesAlongz0} shows $\Psi_2$ and $\Psi_4$ over time along $u=v$ for the different values of $\lambda$. In particular, we see that they do not converge to zero for $u,v>0$ as $a\rightarrow\infty$ as in the case of one wave. This is to be expected from comparison with the Khan-Penrose solution, which already has non-vanishing $\Psi_0,\,\Psi_2$ and $\Psi_4$ in the region after scattering, as well as a theorem by Szekeres \cite{szekeres1965gravitational}.

\begin{figure}[H]
    \centering
    \includegraphics[width=0.5\linewidth]{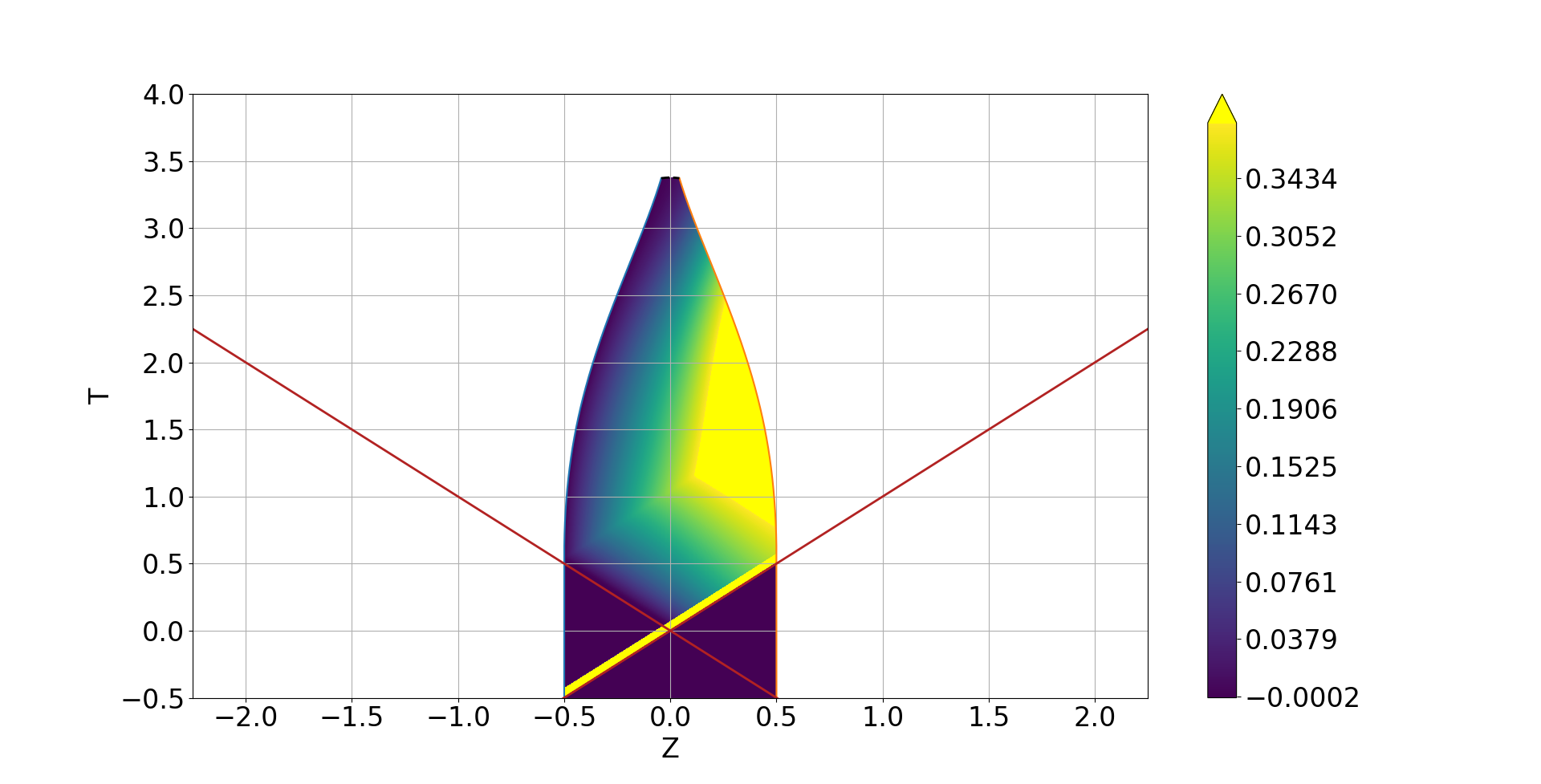}
    \caption{A contour plot of $\Psi_4$ for $a=128$ and $\lambda=0.72$ in the case of colliding impulsive waves.}
    \label{fig:Psi4ContourCollidingWaves}
\end{figure}

Of particular note is the abrupt change in sign of the first time derivative of $\Psi_4$ for $\lambda=0.72$. This sharp turn, which is smooth with a small enough timestep, does not appear this distinctly in any other system variables, except for $\Psi_0$ due to symmetry. Fig.~\ref{fig:Psi4ContourCollidingWaves} shows that this turning point occurs not only at some point along $u=v$ but along an entire null surface which follows the characteristic of $\Psi_4$ from the point where the left boundary hits $v=0$. This is the result of the vanishing boundary condition for $\Psi_4$ on the left boundary being in disagreement with the non-vanishing $\Psi_4$ tail generated by $\Psi_0$ as it passes through the boundary. While at first sight it makes sense to impose a no ingoing radiation condition, this is blatantly unphysical when the evolution itself creates ingoing modes. Between the boundary condition and the evolution equation it is the latter which is fundamental. The boundary condition is nearly completely free to choose and is put in ``by hand''. A common question in a non-linear regime with boundaries that contain both ingoing and outgoing modes is then: How does one make consistent the ``corner condition'', i.e. the physical compatibility between data on a timeslice induced via evolution and the boundary data to yield a physically meaningful result? The answer is simply that there is no clear way to prescribe boundary conditions that match the values in the interior unless one already has an exact solution.


\section{Suppressing the expansion with a train of waves}\label{sec:SupressingExpansion}
In \cite{tsamis2014classical}, the rate of change of an expansion rate parameter $H_{TW}$ with respect to time on a space-like initial value surface (IVS) was calculated to be $N(H^2 - (1/3)K^{ab}K_{ab})$, where $N$ is the lapse in their coordinate system, $K_{ab}$ is the extrinsic curvature to the IVS and $H$ is as per our definition of dS in inflationary coordinates. They hypothesize that there should be no reason why initial data cannot be chosen to satisfy $K^{ab}K_{ab} > 3H^2$ so that the expansion is slowed down and even completely halted\footnote{$N$ is a lapse and should always be positive.}. We can investigate this numerically without solving the constraints by simply choosing dS initial data together with a variety of boundary conditions and seeing how the space-time evolves. We thus explore how a train of waves, generated by choosing the boundary conditions for $\Psi_0$ and $\Psi_4$ appropriately, might accomplish this. 

To do so, we fix $\lambda=0.6$ and define a new function
\begin{equation}\label{eq:streambc}
    p_{stream}(x) = \cases
        {
            32a\cos(c\,x^2)^8\sin(b\,x)^8       & $\displaystyle0<x<\sqrt{\frac{\pi}{2c}}$ \cr
            0       & otherwise
        },
\end{equation}
where $a=0.894,\,b=3129\pi/128000,\,c=1/3$ and we choose $\Psi_0(v,1) = p(v),\,\Psi_4(u,-1) = p(u)$. These constants were chosen through trial and error to give the largest decrease in the expansion while maximizing the interval of time this occurred, before either a singularity is formed or the space-time starts to approach dS again. The cosine factor has the effect of decreasing the amplitude of the wave until it completely vanishes at $ct^2=\pi/2$. This is to hold off a future singularity forming, while still decreasing the expansion rate $\mathcal{H}$.

\begin{figure}[H]
    \centering
    \subfloat[$\mathcal{H}$]
    {{\includegraphics[width=0.5\linewidth]{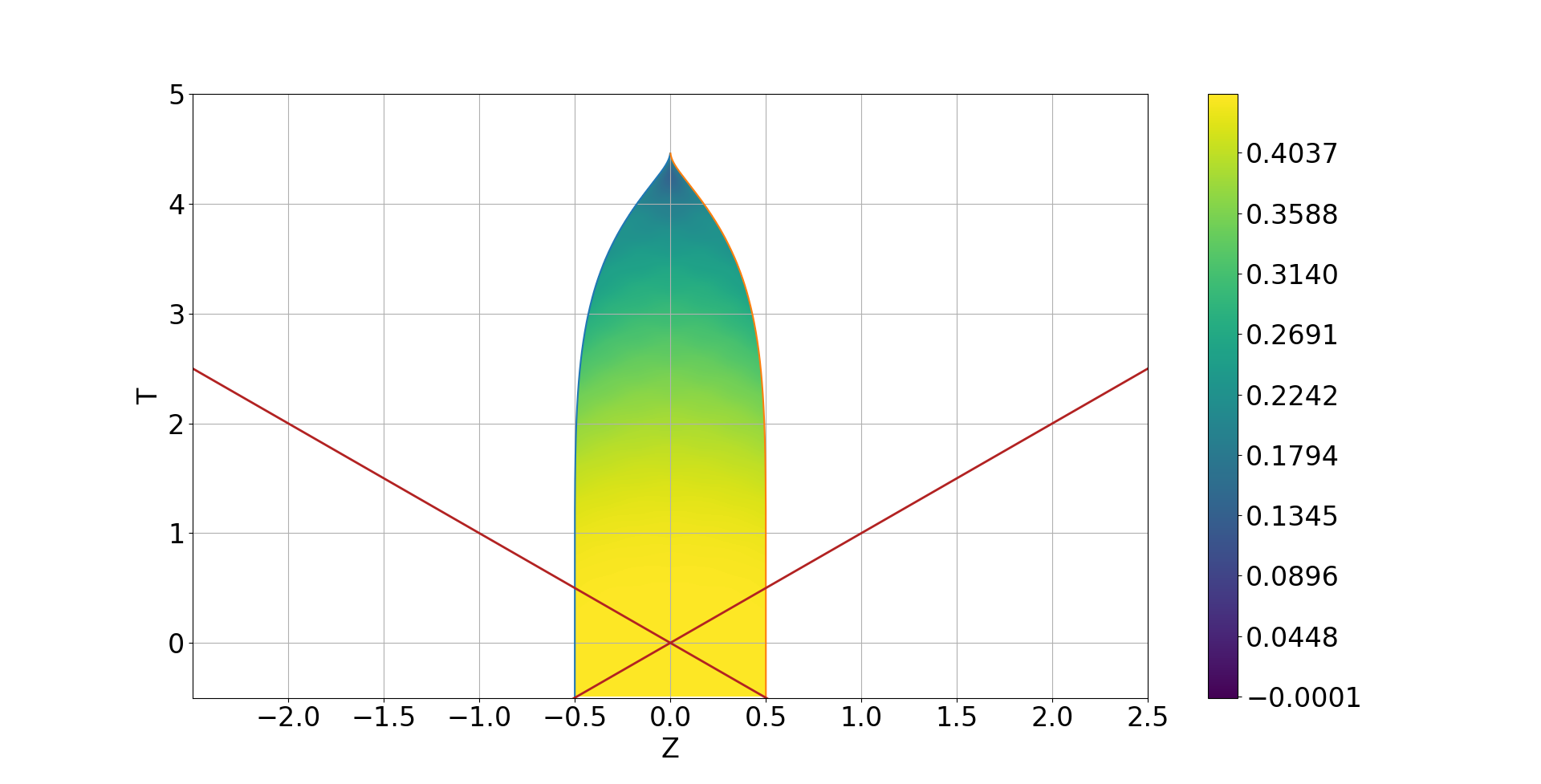}}}
    \qquad
    \subfloat[$I$]
    {{\includegraphics[width=0.5\linewidth]{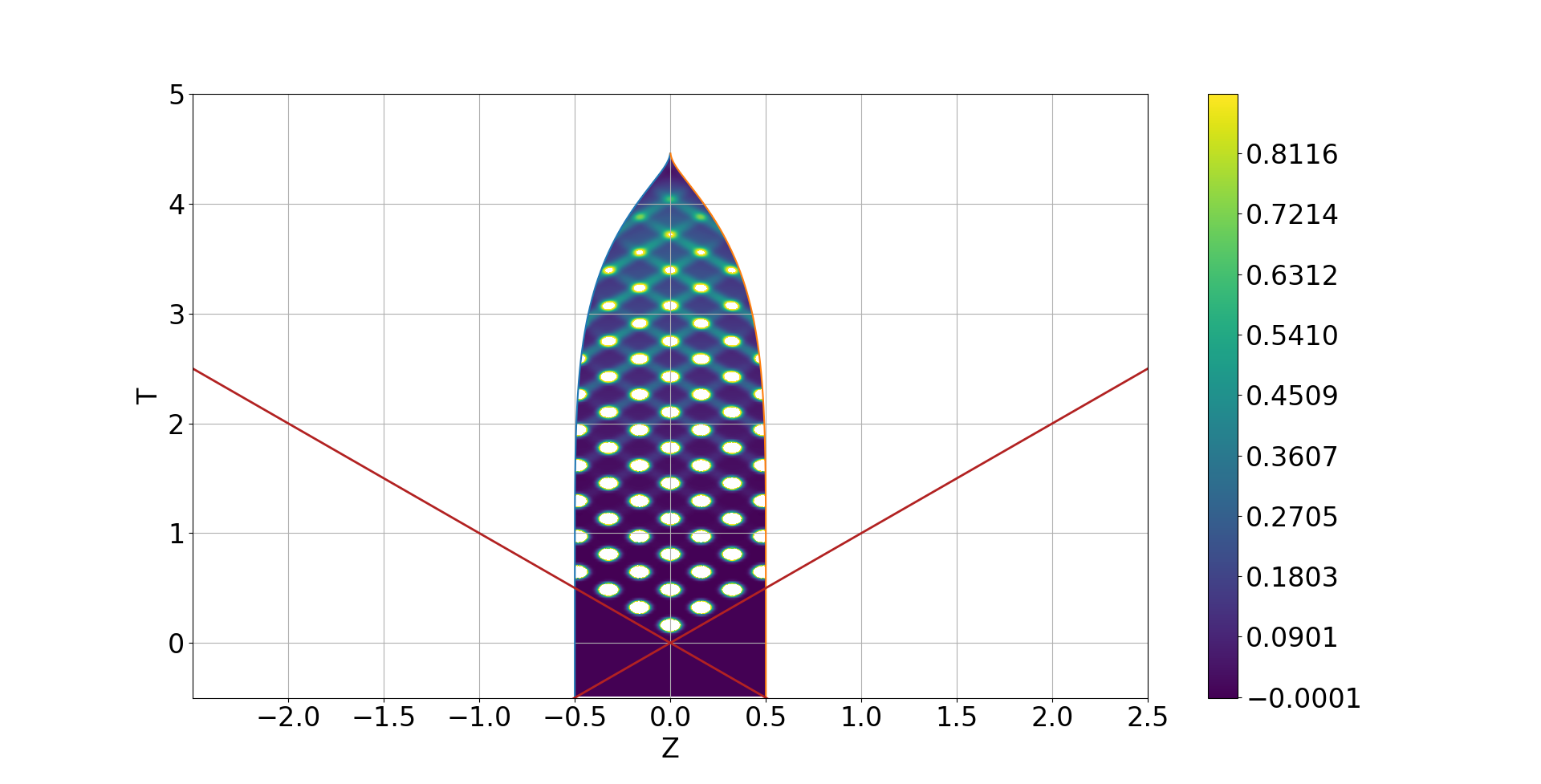}}}
    \caption{Our expansion rate $\mathcal{H}$ and the Weyl invariant $I$ where boundary conditions were chosen using Eq.~\eref{eq:streambc} and with $\lambda=0.6$.}
    \label{fig:HIContourStream}
\end{figure}

Fig.~\ref{fig:HIContourStream} shows the expansion rate $\mathcal{H}$ and Weyl invariant $I$ as contour plots. We see that the expansion rate slowly declines across the entire spatial domain and after a long time ($t\approx14$), forms a coordinate singularity as in case 2. This Weyl invariant shows clearly where the collision regions are, and after a time the waves begin to drag more and more curvature along with them as a tail.

\begin{figure}[H]
    \centering
    \includegraphics[width=0.5\linewidth]{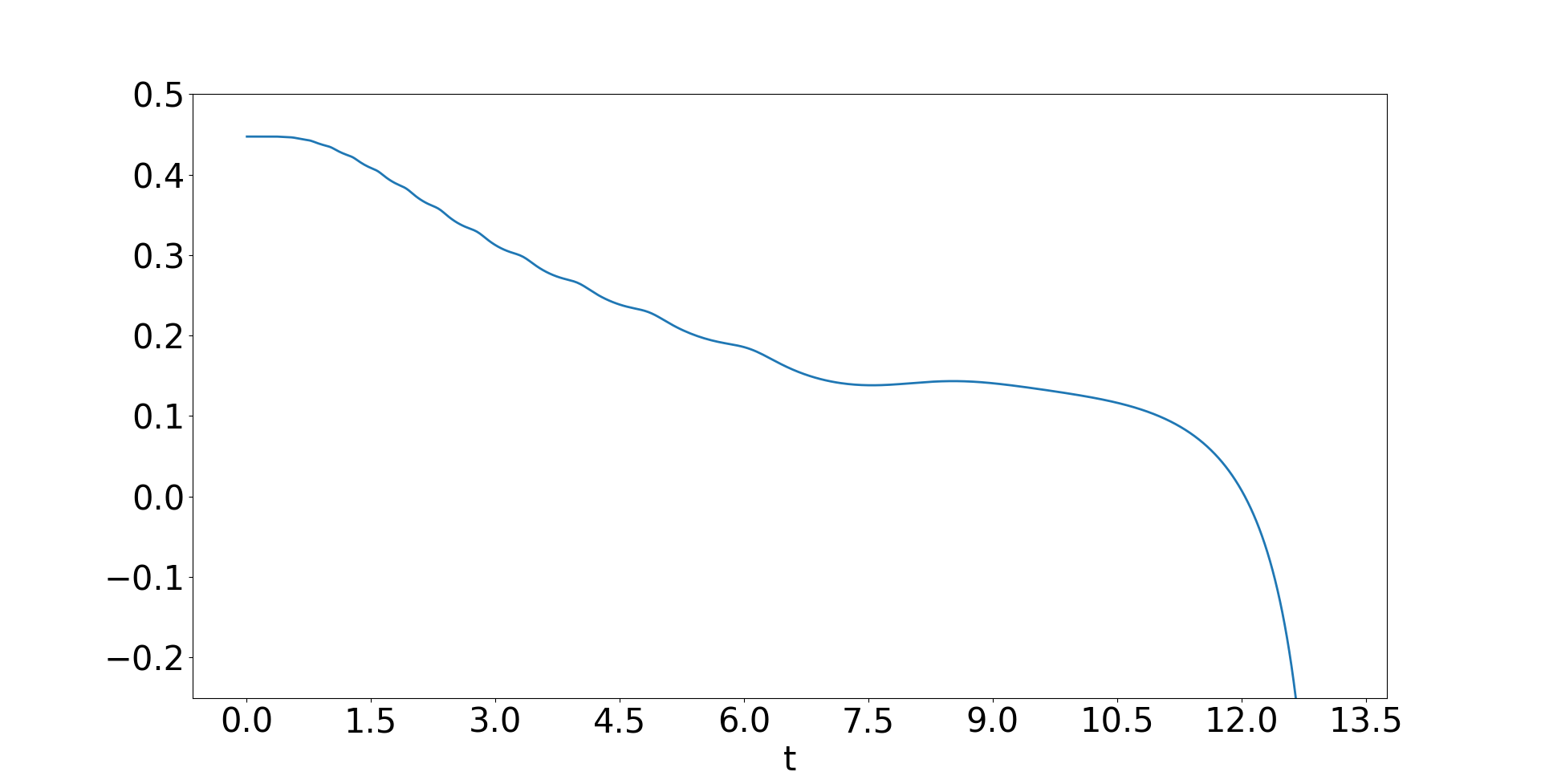}
    \caption{Our expansion rate $\mathcal{H}$ along $z=0$ where boundary conditions were chosen using Eq.~\eref{eq:streambc} and with $\lambda=0.6$.}
    \label{fig:HStreamAlongZ0}
\end{figure}

Fig.~\ref{fig:HStreamAlongZ0} shows just how long we can decrease the expansion for, while holding off forming a singularity. Note that along a spatially constant curve the simulation time $t$ is the proper time of a free falling observer along this curve, and so when we talk about trying to maximize the length of time before a singularity occurs, it is inherently physical. In previous sections we have found that if a singularity was to form after a collision of two waves (with our wave profile), this happens after a few $t$. However here, we can decrease the expansion considerably, without forming a singularity, for up to $t=13$. We cannot however find boundary conditions that lower the expansion rate to zero without very quickly forming a singularity. It is certainly possible that such finely tuned boundary conditions exist, but our studies suggest that they would be very special.

\section{Summary and discussion} \label{sec:summary}

In this paper we have put forward the full non-linear Einstein equations with cosmological constant and non-vanishing energy momentum with the assumption of plane symmetry. These equations were realized through the Newman-Penrose formalism and the imposition of the Friedrich-Nagy gauge, leading to a wellposed initial boundary value problem with timelike boundaries. We specialized to vacuum where $\lambda>0$ and chose initial data to be that induced by the de Sitter space-time in inflationary coordinates. This allowed the exploration of how this space-time is affected by gravitational perturbations, which we generated through appropriate boundary conditions for $\Psi_0$ and $\Psi_4$.

It was found that when only one of the waves was non-vanishing the space-time either wiped out the wave via expansion, or the wave was too strong and a future singularity was produced. The bifurcation was studied and did not produce any critical behaviour. The wave profile was taken to approximate the Dirac delta function to analogize with a known exact solution for $\lambda=0$.

With both waves non-vanishing, and in the physically motivated Gau\ss\; gauge, we found three distinct situations: The waves were not strong enough to cause a contraction of our timelike curves to create a singularity, a coordinate singularity is formed but the curvature remains finite, or a curvature singularity is formed. The second case shows that we can create a singularity where the Weyl invariant $I$ does not diverge, but our expansion parameter diverges to negative infinity along, and close to, the surface $u=v$. The critical behaviour of the two bifurcations separating these futures was explored. Impulsive wave profiles were approximated and it was shown that two bifurcations occur in this case as well.

We encountered two numerical pitfalls during our exploration. Firstly, our free evolution resulted in a false steady state solution close to the bifurcation of the single wave case. As we chose our wave area closer to the bifurcation value, our free evolution approached a steady state (while $A$ was still evolving in time) that did not satisfy the constraints. This happened even though the constraints were satisfied and converged above and below this critical value, showing how careful one must be in monitoring constraints during a free evolution. Secondly, we found that the combination of the evolution system and our non-radiating boundary conditions became unphysical in the colliding wave case after the waves left the computational domain through the boundaries. This was due to the backreaction of the waves creating tails of ingoing radiation, at odds with the boundary conditions. This is however, independent of the fact that our system is wellposed and numerically stable. The question as to how one could ``guess'' the right boundary conditions is delicate and creates a problem that all non-linear simulations, in particular in numerical relativity, face.

We presented how the above situations affected the local expansion rate $\mathcal{H}$, which was taken to be the mean extrinsic curvature of our timeslices up to a constant factor. It was shown that for the case of two waves colliding, we could decrease $\mathcal{H}$ substantially for a long period of time, where the cut-off was determined by numerical limitations, before the space-time asymptoted back to dS. We could do a similar thing with a continuous stream of waves, making the expansion rate drop more uniformly over the computational domain. This showcased the potential to lower the expansion rate over a wider spatial interval.

Although we were not able to find boundary conditions that completely halted expansion for a period before either asymptoting back to de Sitter space-time or forming a singularity, we could still lower it substantially for a long time. Even so, our results do not violate the hypothesis of Woodard and Tsamis', namely that our universe may be in an unstable gravitationally bound state. It would be interesting to see whether, with further testing, we may be able to find boundary conditions that do completely halt expansion for a period.

Now that exploration toward the behaviour of plane gravitational waves with $\lambda>0$ has started and details have been uncovered, it would also be interesting to see whether one can use the results as hints toward an exact solution for impulsive waves. For the case of one propagating impulsive wave, knowing that the Weyl components vanish in the region after the wave has passed should already be a good start.


\section*{References}

\bibliographystyle{elsarticle-num} 
\bibliography{refs}

\end{document}